%% file: hoard_pp.tex
\documentclass[11pt,preprint]{aastex}

\newcommand{\fuse}{{\em FUSE\/ }}
\newcommand{\iue}{{\em IUE\/ }}
\newcommand{\hst}{{\em HST\/ }}

\begin{document}

\title{Observations of the SW Sextantis star DW Ursae Majoris 
with the {\em Far~Ultraviolet Spectroscopic Explorer\/}\footnote{Based 
on observations with the NASA-CNES-CSA {\em Far Ultraviolet Spectroscopic 
Explorer\/}. \fuse is operated for NASA by the Johns Hopkins 
University under NASA contract NAS 5-32985.}}

\author{D.\ W.\ Hoard}
\affil{\vspace*{-11pt}SIRTF Science Center, California Institute of 
Technology, Mail Code 220-6, 1200 E. California Blvd., Pasadena CA 91125}
\email{\vspace*{-11pt}hoard@ipac.caltech.edu}
\authoraddr{SIRTF Science Center, California Institute of Technology, 
Mail Code 220-6, 1200 E. California Blvd., Pasadena CA 91125}

\author{Paula Szkody}
\affil{\vspace*{-11pt}Department of Astronomy, University of Washington, 
Box 351580, Seattle WA 98195-1580} 
\email{\vspace*{-11pt}szkody@astro.washington.edu}
\authoraddr{Department of Astronomy, University of Washington, Box 351580, 
Seattle WA 98195-1580} 

\author{Cynthia S.\ Froning}
\affil{\vspace*{-11pt}Center for Astrophysics and Space Astronomy, 
University of Colorado, 389 UCB, Boulder CO 80309-0389}
\email{\vspace*{-11pt}cfroning@casa.colorado.edu}

\author{Knox S.\ Long}
\affil{\vspace*{-11pt}Space Telescope Science Institute, 
3700 San Martin Drive, Baltimore MD 21218}
\email{\vspace*{-11pt}long@stsci.edu}

\author{Christian Knigge}
\affil{\vspace*{-11pt}Department of Physics and Astronomy, 
University of Southampton, Highfield, Southampton SO17 1BJ, UK}
\email{\vspace*{-11pt}christian@astro.soton.ac.uk}

\slugcomment{To appear in the Astronomical Journal, November 2003}

\begin{abstract}
We present an analysis of the first far-ultraviolet observations 
of the SW Sextantis-type cataclysmic variable DW Ursae Majoris, 
obtained in November 2001 with the {\em Far Ultraviolet 
Spectroscopic Explorer}.  The time-averaged spectrum of DW UMa 
shows a rich assortment of emission lines (plus some contamination 
from interstellar absorption lines including molecular hydrogen).  
Accretion disk model spectra do not provide an adequate fit to the 
far-ultraviolet spectrum of DW UMa.  We constructed a light curve 
by summing far-ultraviolet spectra extracted in 60-sec bins; this 
shows a modulation on the orbital period, with a maximum near 
photometric phase 0.93 and a minimum half an orbit later.  No other 
periodic variability was found in the light curve data.  We also 
extracted spectra in bins spanning $\Delta\phi=0.1$ in orbital phase; 
these show substantial variation in the profile shapes and velocity 
shifts of the emission lines during an orbital cycle of DW UMa.  
Finally, we discuss possible physical models that can qualitatively 
account for the observed far-ultraviolet behavior of DW UMa, in the 
context of recent observational evidence for the presence of a 
self-occulting disk in DW UMa and the possibility that the SW Sex 
stars may be the intermediate polars with the highest mass transfer 
rates and/or weakest magnetic fields.
\end{abstract}


\section{Introduction}
\label{s-intro}

Cataclysmic variables (CVs) are semi-detached binary stars in which 
a late main sequence star loses mass onto a white dwarf (WD) via 
Roche lobe overflow.  In systems containing a non-magnetic WD, 
accretion proceeds through a viscous disk.  
In systems containing a 
magnetic WD, accretion may proceed directly from the L1 point onto 
the field lines in the strong field case, or through a partial disk 
in the case of intermediate field strength.
The former systems are known as ``polars'' (owing to their highly 
polarized radiation), while the latter are ``intermediate polars'' (IPs). 
In polars, the strong WD magnetic field prevents the formation of a disk.
In IPs, the field truncates only the inner edge of the disk, and
material is accreted onto the WD through two
accretion ``curtains'' (see review by \citealt{patterson94}). 
See \citet{warner95} for a thorough review of CV types and behavior.

Over a decade ago, three CVs (SW Sextantis, DW Ursae Majoris, and 
V1315 Aquilae) were proposed as the founding members of a new CV 
subclass, the ``SW Sex stars'' \citep{honey86,szkody90,thor91}.  
Additional CVs matching the characteristics of the first 
three were subsequently identified.  
The original defining properties of the SW Sex stars are as 
follows (also see \citealt{thor91}, \citealt{hoard98}, and the 
review by \citealt{warner95}):

\begin{itemize}
\item{They are ``novalike'' CVs.  Unlike the more well-known dwarf 
nova class of CV, members of the novalike class do not undergo 
quasiperiodic outbursts.  Novalikes are instead characterized by 
an approximately steady, high rate of mass transfer (and 
correspondingly prominent accretion disk) that quenches the disk 
instability mechanism responsible for dwarf nova outbursts.}

\item{Their optical light curves show deep eclipses of the WD+disk by 
the secondary star, requiring the system inclination to be 
high ($i\gtrsim80^{\circ}$).}

\item{They have orbital periods of 3--4 hr, just above the 2--3 hr 
``period gap'' (in which few non-magnetic CVs are found) that is 
thought to arise during the secular evolution of CVs due to a 
change in angular momentum loss mechanism, likely triggered by 
mass-loss-induced evolution of the donor star (e.g., 
\citealt{shafter92}, \citealt{howell01}).  
This range of orbital periods is also associated with the 
largest range and highest extreme of predicted 
rates of mass transfer in CVs \citep{wu95}, above expectations 
from the standard theory for the secular evolution of CVs.}

\item{They display high levels of spectral excitation, including 
\ion{He}{2} $\lambda4686$ emission that is often comparable in 
strength to H$\beta$.}

\item{Their spectra exhibit single-peaked emission lines rather than 
the double-peaked lines expected from near-edge-on disks.}

\item{The Balmer and \ion{He}{1} emission lines are only shallowly 
(or not at all) eclipsed compared to the continuum (implying 
emission originating above the orbital plane).}

\item{The zero-crossings of their emission 
line radial velocities exhibit pronounced phase offsets relative to 
their eclipse ephemerides (implying a 
non-uniform distribution of emitting regions in the disk).}

\item{Transient absorption features appear in their Balmer 
and \ion{He}{1} emission line cores, typically around photometric 
phase $\phi=0.5$ (i.e, the superior conjunction of the secondary star, 
opposite the eclipse).}
\end{itemize}

The number of confirmed and probable SW Sex stars has now swelled 
to $\approx20$ systems\footnote{See ``The Big List of SW Sextantis Stars'' 
at {\url http://spider.ipac.caltech.edu/staff/hoard/research/swsex/biglist.html}.}, 
including more than half (16 out of 30) of all securely identified 
novalike CVs with $P_{\rm orb}=3$--4 hr.
In part, this is due to accepting that ``SW-Sexiness'' does not 
have a rigid definition linked to a specific CV morphology.  
Rather, the original SW Sex stars represent extreme cases of a 
pathology that is present at some level in many
high mass transfer rate CVs.  
Consequently, SW Sex behavior has been recognized 
in CVs that do not eclipse (e.g., V442 Ophiuchi -- \citealt{hoard00}), 
have orbital periods outside the range 3--4 hr 
(e.g., BT Monocerotis -- \citealt{smith98}), or in which the 
transient absorption occurs outside the orbital phase range 
$\phi\approx0.4$--0.6 (e.g., UU Aquarii -- \citealt{hoardetal98}).  

DW UMa is one of the three archetype SW Sex stars.  It was first 
studied in detail by \citet[][also see \citealt{zhang88}]{shafter88}, 
who obtained extensive photometric and spectroscopic data.  
Optical light curves of DW UMa show a deep, V-shaped 
asymmetric eclipse, and short time scale variability with amplitude 
$\sim0.1$ mag at orbital phases away from eclipse.  
These light curves do not contain a pre-eclipse hump (caused by a 
bright spot at the impact site of the accretion stream from the secondary star 
with the outer edge of the disk).  
The optical spectrum of DW UMa shows strong, 
single-peaked emission lines of H and He (I and II), as well as 
\ion{C}{2} $\lambda4267$ and the 
\ion{C}{3}/\ion{N}{3} $\lambda4645$ Bowen blend. 
There is no evidence of a bright spot at optical wavelengths in either 
the light curve of DW UMa or more detailed studies based on 
Doppler tomography \citep{kaitchuck94,hoard98}.

Although the mass transfer process in novalike CVs is (on average) 
steady, some novalikes (the so-called VY Sculptoris stars) 
occasionally undergo low 
states in which mass transfer is severely reduced or shut 
off completely.  When this happens, the accretion disk 
decreases in size or disappears completely, and the overall 
system brightness fades by several magnitudes.  
When DW UMa was observed with the Space Telescope Imaging 
Spectrograph (STIS) on the {\em Hubble Space Telescope (HST)\/} 
in January 1999, it had just dropped into such a low state.  
This allowed \citet{knigge00} to determine, via comparison with 
normal -- or ``high'' -- state spectra obtained with the 
{\em International Ultraviolet Explorer (IUE)\/}, that the 
accretion disk in DW UMa completely obscures the WD during the 
high accretion state.  They concluded that the disk in DW UMa increases
in thickness from near the WD to its outer edge (i.e., it is 
a ``flared'' disk), and the optically thick outer rim of the disk is 
responsible for the self-occultation of the WD and inner disk.
Recently, \citet{araujo03} used these 
same \hst data to determine reliable system parameters for DW UMa, 
including a mass ratio of $q = M_{2}/M_{\rm WD} = 0.4\pm0.1$ 
(with $M_{\rm WD}/M_{\odot} = 0.77\pm0.07$), inclination of 
$i = 82^{\circ} \pm 4^{\circ}$, orbital separation of 
$a/R_{\odot} = 1.14\pm0.06$, and distance of $d = 600$--900 pc.  
In this paper, we present an 
analysis of the first far-ultraviolet (FUV) observations of 
DW UMa, obtained with the {\em Far Ultraviolet Spectroscopic 
Explorer (FUSE)\/}.

\section{Observations}
\label{s-mainobs}

\begin{figure}[tb]
\epsscale{0.94}
\plotone{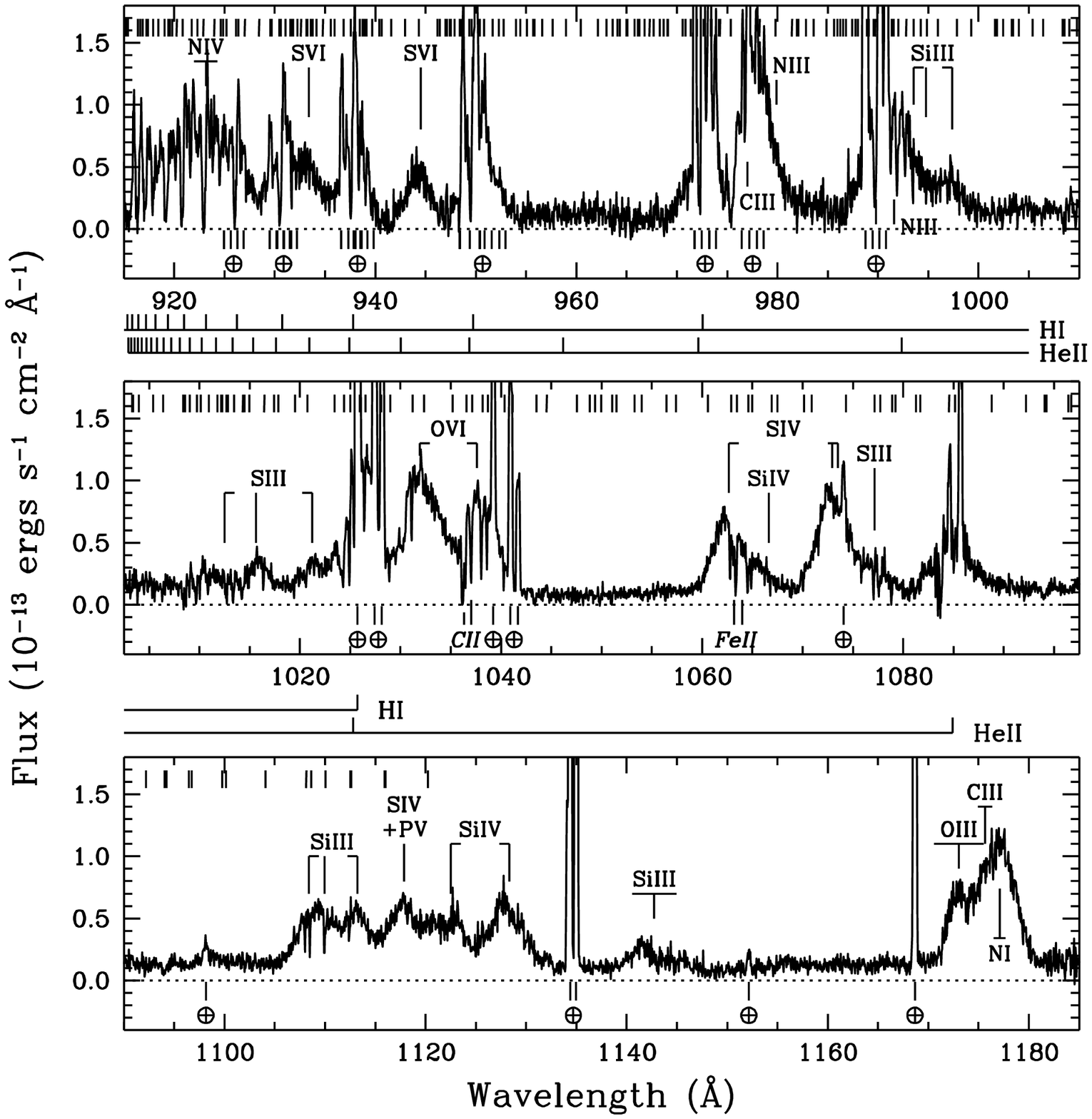}
\epsscale{1.00}
\figcaption{Total combined FUV spectrum of DW UMa.  
The spectrum spans 915--1185 \AA; the top and bottom panels 
overlap the ends of the middle panel by 7.5 \AA.  
Emission lines are labeled, with widely spaced lines of 
the same multiplet indicated by short horizontal bars on 
the ``bookend'' transitions, and multiplets with many 
closely spaced lines indicated by a vertical pointer at 
the midpoint wavelength joined to a horizontal bar 
spanning the multiplet transition wavelengths.  
Airglow lines ($\oplus$; identified from \citealt{feldman01}) 
and ISM lines (ions in italics) are indicated below the spectrum.  
The airglow lines have been truncated at the upper flux limit of the plot.
Wavelengths of \ion{H}{1} and \ion{He}{2} transitions are 
shown outside and below the top and middle panels.  
Unlabeled hashmarks at the top of each panel are 
interstellar molecular hydrogen (H$_{2}$) transitions 
(see Section \ref{s-h2}).  
\label{f-fuvspec}}
\end{figure}

We observed DW UMa with \fuse during 6 exposures (separated by 
Earth occultations) between 
HJD 2452221.26--2452221.62 (2001 November 07, 18:14 -- November 08, 02:43 UT; 
see Table \ref{t-fuselog}).  
All data were obtained using the LWRS aperture and TTAG accumulation 
mode (for \fuse spacecraft and instrument details see, 
for example, \citealt{sahnow00}\footnote{Also see the \fuse Science Center 
web page at \url{http://fuse.pha.jhu.edu/}.}).
We used the CalFuse v2.2.3 pipeline software to extract the FUV spectra 
from the raw data files obtained during each \fuse exposure.
We then used a custom-built IDL routine (following the recipes in 
the \fuse Data Analysis Cookbook\footnote{See 
\url{http://fuse.pha.jhu.edu/analysis/cookbook.html}.}) to combine 
the various detector and mirror segments of the spectra.  
This yields a time-averaged spectrum with a total equivalent 
exposure time of $\approx 18.8$ kiloseconds.
We excluded the LiF1 data for $\lambda = 1117$--1149 \AA\ in 
order to avoid the artifact known as ``The Worm'' (only the LiF2 data 
in this region are used).  The final combined spectrum 
(see Figure \ref{f-fuvspec}) was rebinned 
onto a uniform wavelength scale with dispersion 0.05 \AA\ pixel$^{-1}$
by averaging flux points from the original dispersion 
(0.007 \AA\ pixel$^{-1}$) into wavelength bins of width 0.05 \AA.

\begin{figure}[tb]
\epsscale{0.94}
\plotone{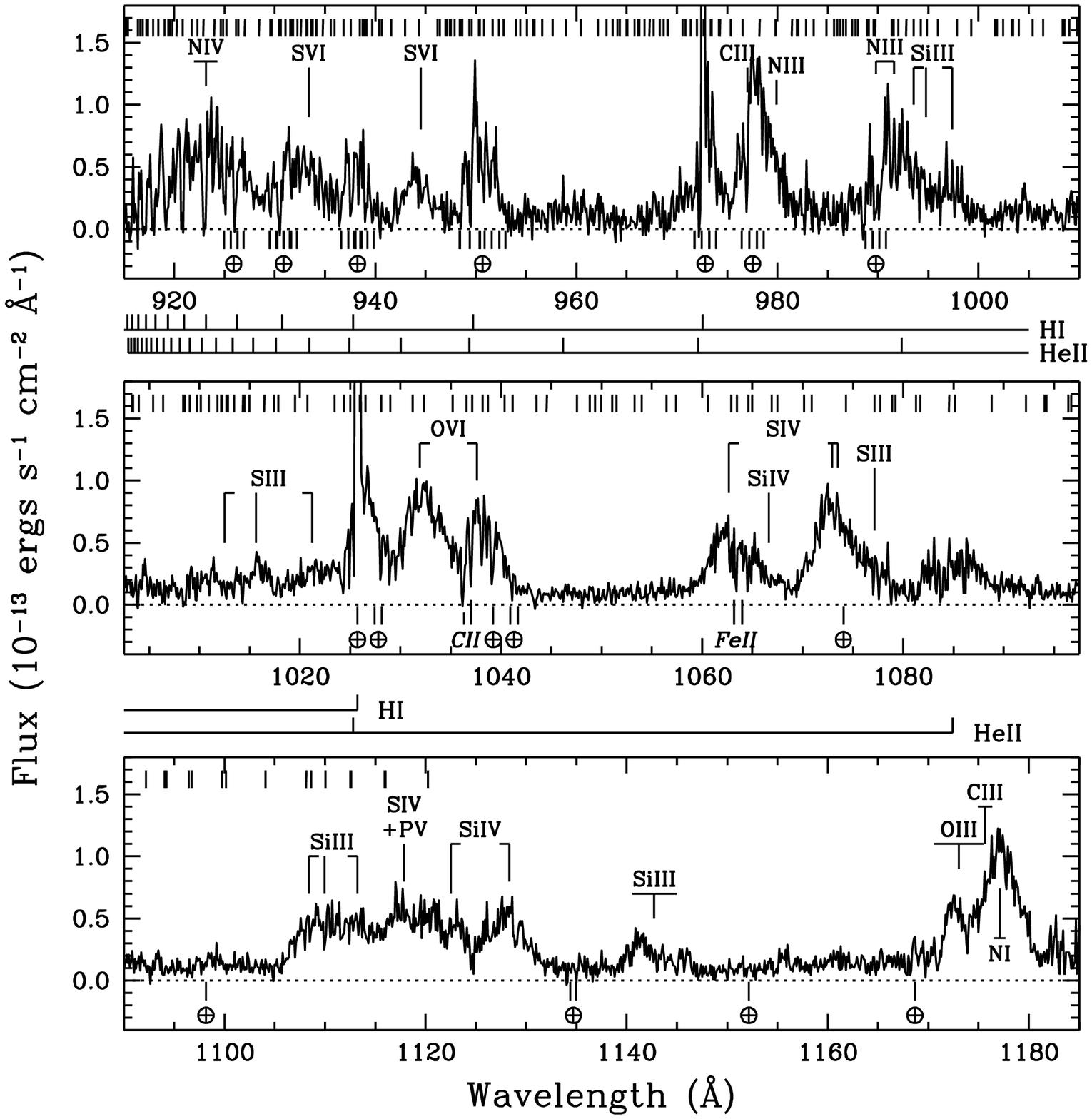}
\epsscale{1.00}
\figcaption{As in Figure \ref{f-fuvspec}, but showing only 
the data obtained during satellite night.  
Note the absence of strong airglow features (with the 
exception of Ly-$\beta$) at the wavelengths indicated with 
a $\oplus$ symbol.
\label{f-fuvspecnight}}
\end{figure}

We also repeated the spectrum extraction as described above, 
but using custom screening files that exclude data obtained 
during satellite day (when airglow contamination is strongest).  
This results in a total effective exposure time of only 
$\approx 4$ kiloseconds (about 22\% of the total day+night exposure).  
The night spectrum (see Figure \ref{f-fuvspecnight}) was rebinned 
to a dispersion of 0.10 \AA\ pixel$^{-1}$ in order to more closely 
match the signal-to-noise ratio (S/N) 
obtained in the longer day+night spectrum.  
Comparison between the two spectra clearly demonstrates which features 
are artifacts due to airglow. The night spectrum also reveals some DW UMa 
emission features that were effectively masked by airglow in the 
day+night spectrum (especially at the short wavelength end of the 
spectrum).  Meanwhile, the much longer exposure time of the day+night 
spectrum allows us to resolve finer detail in the profiles of the 
emission lines (e.g., in the emission complex between 1105--1130 \AA).

The time-averaged (day+night) \fuse spectrum of DW UMa has a 
mean continuum level of 
$F_{\lambda}\approx1.2\times10^{-14}$ erg s$^{-1}$ cm$^{-2}$ \AA$^{-1}$, 
with a sharp cut-off at the Lyman limit.
A number of narrow interstellar absorption lines of various metal ions, 
\ion{H}{1}, and molecular hydrogen (see Section \ref{s-h2}) 
are visible in the spectrum.
Although contemporaneous photometric data are not available, the 
presence of this wide array of strong emission features in the 
FUV spectrum confirms
that DW UMa was in the normal, high accretion state during 
our \fuse observations.
Some lines that are prominently visible in the spectrum 
include:\ numerous \ion{H}{1} (including Lyman-$\beta$) and 
\ion{He}{1} lines, \ion{O}{6} $\lambda\lambda1031.9, 1037.6$, 
the \ion{C}{3} multiplet at 1175.3 \AA, and many \ion{S}{3}, 
\ion{S}{4}, \ion{Si}{3}, and \ion{Si}{4} multiplets.
These emission lines have complex profiles in which, often, several ionic 
species and/or multiplet transitions are blended together; the additional
presence of airglow (emission) and interstellar (absorption) 
features further complicates many of the line profiles.  

In Table \ref{t-lines}, we have listed the equivalent widths (EWs) for 
several lines in the FUV spectrum of DW UMa (measured by direct 
integration of pixel values between manually selected continuum points 
using the ``e'' routine in the IRAF\footnote{The Image Reduction and 
Analysis Facility is distributed by the National Optical Astronomy 
Observatories, which are operated by the Association of Universities 
for Research in Astronomy, Inc., under cooperative agreement with 
the National Science Foundation.} task {\sc splot}).  
Except where noted in the table as approximate values, 
the EWs are reproducible to within 10\%.
Although these lines were selected to be comparatively isolated and/or 
resolved multiplet members, the presence of superimposed interstellar 
and/or airglow features makes the precise values of our directly 
measured EWs somewhat uncertain.  
For comparison, we have also provided the results of Gaussian fits to 
the emission line profiles (using the ``k'' routine in {\sc splot}).  
The closely-spaced Lyman-$\beta$ and \ion{O}{6} doublet lines were fit 
simultaneously (using the ``d'' routine in {\sc splot}), as was 
the \ion{S}{3} multiplet.  
The night-only spectrum was used for the Lyman-$\beta$ and \ion{O}{6} 
measurements to reduce airglow contamination (the residual 
Lyman-$\beta$ airglow in the night-only spectrum was masked off).  
EWs for a few interstellar \ion{H}{1} lines in regions with reliable adjacent
continuum levels are also listed in Table \ref{t-lines}.

\section{Analysis}

\subsection{Molecular Hydrogen Absorption}
\label{s-h2}

The wavelength range over which \fuse observes is susceptible to 
contamination by absorption features of the ubiquitous interstellar 
molecular hydrogen in the lines of sight to targets of 
interest\footnote{In fairness, we note that to some observers, 
these molecular hydrogen lines are the interesting features in 
a \fuse spectrum.}.
We used the ``H2tautemplates'' from the {\sc H2ools} package 
\citep{mccandliss03}\footnote{Also 
see \url{http://www.pha.jhu.edu/$\sim$stephan/h2ools2.html}.} 
to model the effect of interstellar molecular hydrogen absorption 
in our \fuse spectrum of DW UMa.  
The H2tautemplates tabulate molecular hydrogen optical depths as 
a function of wavelength (on a 0.01 \AA\ grid), computed for a 
maximum column density of $\log N_{\rm H_2} = 21$, that can be used to 
calculate normalized template spectra of the H$_{2}$ absorption lines.
Rotational states from j=0--15 are included (the vibrational state 
is fixed at 0).  We considered a maximum rotational state of j=4 
(rotational states above j=4 are difficult to observe in typical 
interstellar clouds; \citealt{shull00}).
The calculated template spectra assume a uniform continuum level 
of 1.0, so they must be scaled to the mean flux level in the 
observed spectrum.  This is problematic, however, because of 
the undulations in the observed \fuse spectrum of DW UMa caused 
by intrinsic continuum variation, as well as the presence of 
emission lines (some of which have superimposed H$_{2}$ 
absorption lines).  To account for this, we fit a high order 
spline function to the day+night \fuse spectrum (with the 
airglow and ISM lines masked off) in the wavelength region 
1000--1185 \AA\ (i.e., the higher S/N LiF data region).  
We then normalized this fit to a mean level of 1.0 in the 
continuum regions and multiplied it with the H$_{2}$ template 
spectra in order to approximate the changing flux level of the 
observed spectrum at different wavelengths.  In addition, we 
convolved the templates with a Gaussian kernel with FWHM 
of 0.06 \AA\ to simulate the instrumental resolution of 
the \fuse spectrum (this value is in agreement with both the 
measured resolving power of {\em FUSE}\footnote{From the 
\fuse Observer's Guide; 
see \url{http://fuse.pha.jhu.edu/support/guide/obsguide.html}.} 
and the widths of the non-H$_{2}$ ISM lines in our spectrum). 

We determined the best fitting H$_{2}$ absorption model by 
manually adjusting the parameters that affect the calculation 
of the template spectra (scaling factor, radial velocity shift, 
column density, Doppler parameter\footnote{The Doppler 
parameter is the sum -- added in quadrature -- of the 
thermal and turbulent velocities in the absorbing cloud.  
The threshold column density at which a given line begins 
to saturate will be lower (higher) for a low (high) Doppler 
parameter.}) until a minimum r.m.s.\ 
deviation between the template and data, in selected regions 
containing H$_{2}$ lines, was reached.
The best value of each parameter can be determined essentially 
independent of the other parameters by considering different 
aspects of the observed spectrum whose behaviors are most 
affected by a given parameter.
First, the overall level of the template spectrum is scaled 
to match the mean continuum level of the observed spectrum.  
As used here, this scaling factor is physically meaningless, 
and only serves to match the flux scales of the template and 
observed spectra.
The radial velocity shift of the absorbing cloud can be 
determined even before reasonable values of the other 
parameters have been estimated by simply comparing the 
centers of the observed H$_{2}$ absorption lines with those in the template.
The molecular hydrogen column density must be made large 
enough that the weakest H$_{2}$ features are present in 
the template.  The Doppler parameter
can then be adjusted until the saturation level 
of the strongest lines in the template matches the observed spectrum.

\begin{figure}[tb]
\epsscale{1.00}
\plotone{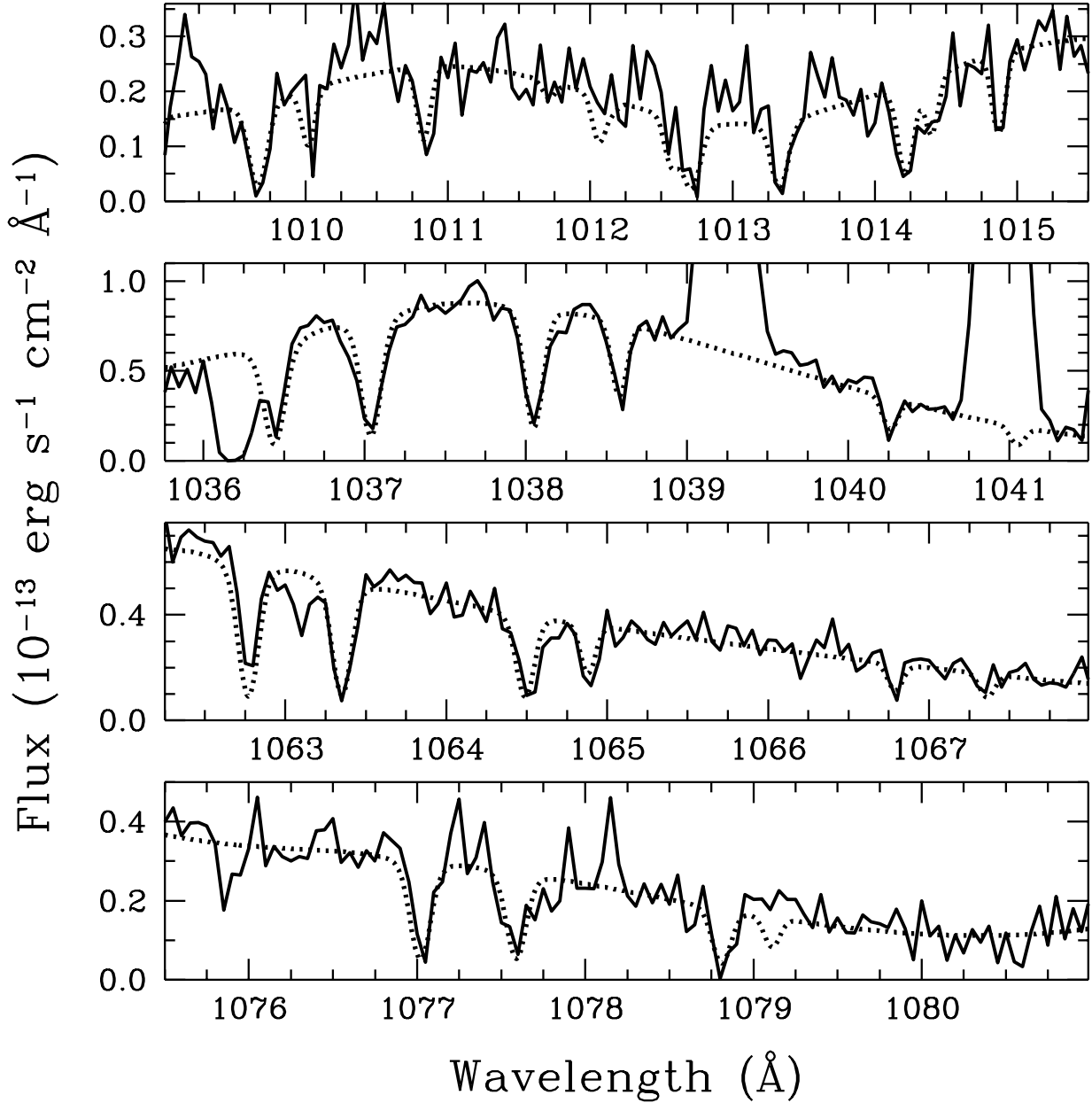}
\epsscale{1.00}
\figcaption{FUV spectrum of DW UMa (solid line) with the 
fitted molecular hydrogen absorption template spectrum 
discussed in the text (dotted line), shown in several 
representative wavelength regions.
\label{f-h2fit}}
\end{figure}

In this manner, we found that the best fitting molecular 
hydrogen absorption model has a column density (molecules cm$^{-2}$) of 
$\log N_{\rm H_2} = 18.5$ $(+0.2/-0.5)$, a radial velocity shift  
of $\Delta v = -33$ $(\pm1)$ km s$^{-1}$, and a Doppler parameter 
of $5$ $(\pm1)$ km s$^{-1}$.
Figure \ref{f-h2fit} shows representative sections of the 
best-fitting template spectrum superimposed on the day+night 
\fuse spectrum of DW UMa.
The $H_{2}$ column density of our model is a moderate value relative to
the range seen in surveys of the Galactic interstellar molecular 
hydrogen content ($\log N_{\rm H_2} \approx 13$--21; 
\citealt{savage77}, \citealt{shull00}, and \citealt{richter03}).
Besides the interstellar \ion{H}{1}, \ion{C}{2}, and \ion{Fe}{2} lines 
indicated in our \fuse spectrum, almost all of the other 
narrow absorption features can be accounted for 
by interstellar molecular hydrogen absorption.  
In principle, it is possible that multiple discrete molecular 
hydrogen clouds along the line-of-sight will all contribute 
absorption with possibly different parameters of column density, 
velocity shift, etc.  However, we see no compelling need for 
more than a single absorbing cloud to adequately (and 
conservatively) reproduce the H$_{2}$ 
absorption lines present in the \fuse spectrum of DW UMa.

\subsection{Near- and Far-Ultraviolet Spectrum}

\begin{figure}[tb]
\epsscale{1.00}
\plotone{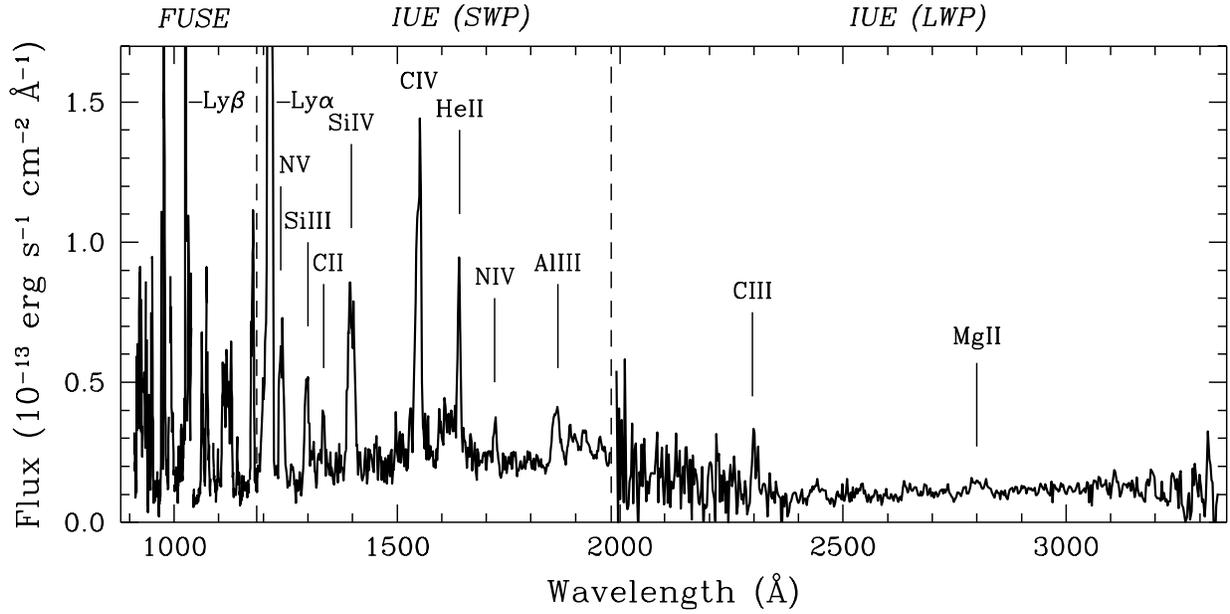}
\epsscale{1.00}
\figcaption{Combined \fuse and \iue spectrum of DW UMa.  
Vertical dashed lines mark the boundaries between the {\em FUSE},
\iue SWP, and \iue LWP wavelength regions.  Several 
features in the \iue sections of the spectrum are identified.  
The strong Lyman-$\alpha$ and 
-$\beta$ features have been truncated, and other strong 
airglow features in the \fuse spectrum have been masked off entirely.
\label{f-iuespec}}
\end{figure}

DW UMa was observed on two occasions with {\em IUE}, during 
November 1985 and February 1987 \citep{szkody87}.  
We retrieved the available spectra from the \iue archive, 
and have plotted them 
together with our \fuse spectrum in Figure \ref{f-iuespec}.  
The \iue short wavelength spectrum is SWP27097 (4500 sec 
obtained on 1985 Nov 13), while the \iue long wavelength 
spectrum is the mean of LWP07086 and LWP07087 (2100 sec and 
2400 s, respectively, obtained on 1985 Nov 13).
The flux levels at the long and short wavelength ends of 
the \fuse and \iue spectra, respectively, match quite closely.  
Overall, DW UMa (in the high accretion state) is revealed to 
have a near-to-far UV (900--3300 \AA) spectrum dominated by 
strong emission lines, and a continuum with a ``bump'' 
between 1300--2300 \AA.
As mentioned in Section \ref{s-intro}, DW UMa was also 
observed in the UV using STIS+\hst \citep{knigge00}; 
however, those data were obtained during an extreme low 
accretion state of the CV, so are not directly comparable 
to our high state \fuse spectrum.

\subsubsection{Attempted Disk Spectrum Modeling}

We attempted to fit various steady-state accretion disk model spectra to 
the combined \fuse + \iue spectrum of DW UMa using the methods 
described by \citet{froning01} and the accretion disk models 
used by \citet{long94}.  We masked off the emission line regions 
in the UV spectrum in order to include only the continuum 
regions in the model fits.  We used the system parameters 
from \citet{araujo03} to initialize the models; however, 
because the flared disk in DW UMa is expected to occult the WD 
in the high accretion state \citep{knigge00}, we did not 
include a contribution from a WD in our model spectra.  
None of our attempts at modeling the UV spectrum of DW UMa 
yielded fits that adequately reproduced the observed continuum shape.  
In all cases, the flux in the disk models increases more rapidly 
to the blue than is observed in the spectrum of DW UMa, 
and the observed 
bump (described above) is not recreated in the models.

\citet{knigge00} have shown that the disk of DW UMa in the 
high accretion state has a flared edge that completely obscures 
the WD.  This would effectively ``remove'' the inner disk 
region from the FUV spectrum by blocking our view of it, and 
produce a flatter-than-expected continuum that is dominated 
by the cool disk rim.  
A similar situation was found during attempts to fit accretion 
disk models to the \fuse + \iue spectrum of the eclipsing novalike CV, 
UX Ursae Majoris \citep{froning03b}.  In that case also, the 
disk model that best fits the FUV spectrum increases in flux 
too rapidly toward its blue end to match the observed data,
although the discrepancy is smaller than we found 
for DW UMa \citep{froning03a}.  
\citet*[][and references therein]{froning03b} suggest several 
possible factors that might contribute to this effect, 
including truncation of the inner disk (which would remove 
flux from the bluest wavelengths), or the presence of an 
additional source of continuum flux (such as an optically 
thin accretion disk chromosphere).
We note that obscuration by 
a thick accretion disk rim might also suffice; 
however, the eclipses in UX UMa are U-shaped \citep{baptista95}, 
whereas a flared disk would be expected to produce eclipses that 
are more V-shaped \citep{knigge00}, as seen in DW UMa and other 
SW Sex stars.  Consequently, one of the other mechanisms suggested 
by \citet*{froning03b} may be more important in UX UMa.
Our collaboration has recently received 
approval for new observations of DW UMa using STIS+\hst that 
will take place during Cycle 12; these should lead to a better 
understanding of the disk geometry
by providing new, high S/N, time-resolved, high accretion 
state UV spectra (which can be directly compared to the STIS 
low state spectra and combined with the \fuse data to extend 
the UV coverage to longer wavelengths for additional modeling attempts).

\subsection{FUV Light Curves}

\begin{figure}[tb]
\epsscale{0.95}
\plotone{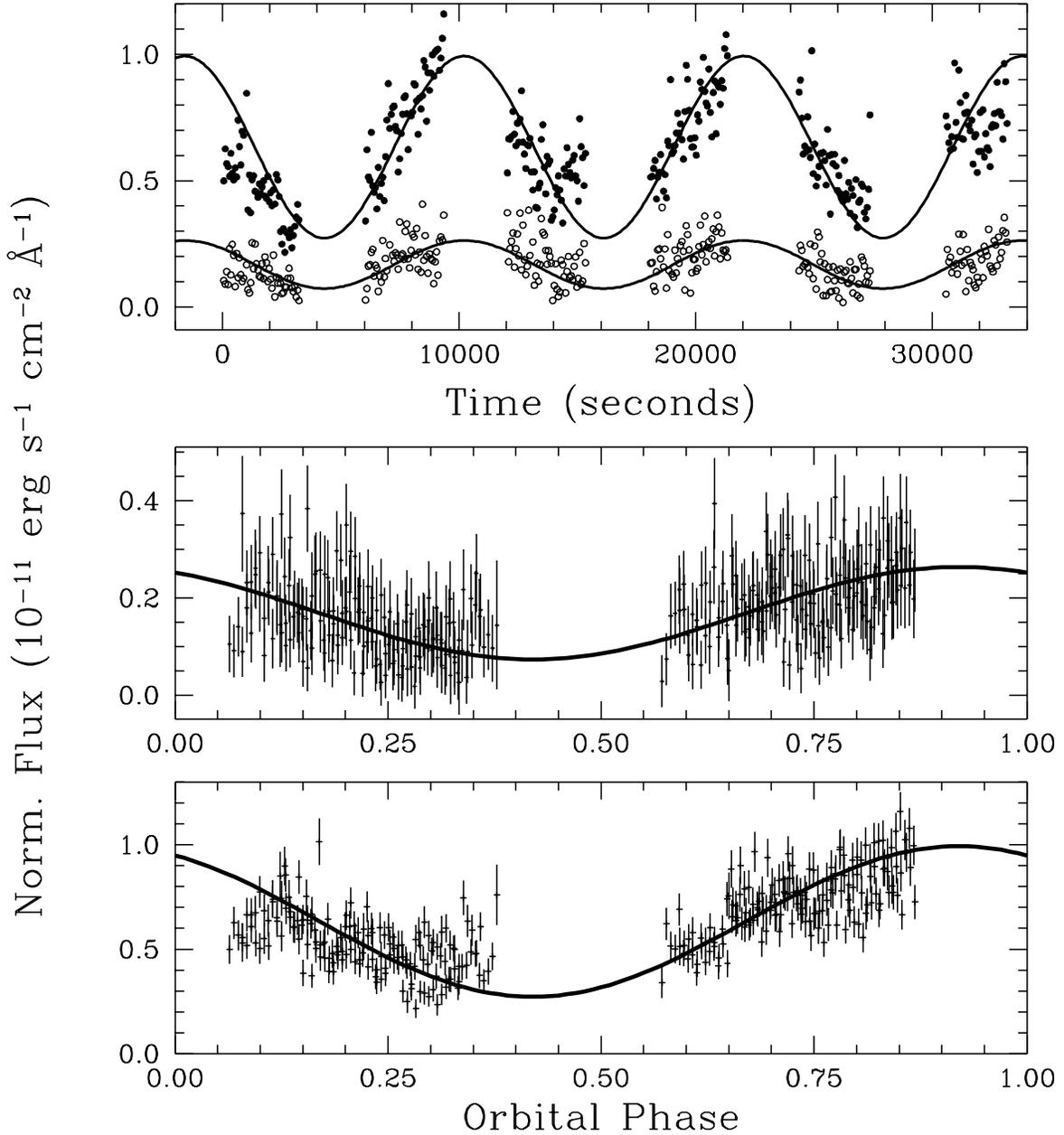}
\epsscale{1.00}
\figcaption{FUV light curves of summed continuum (1154--1167 \AA; 
open circles) and emission line (Si+S, 1105--1131 \AA; filled 
circles) regions in the 60-sec LiF2 spectra of DW UMa (top panel).  
The continuum and 
emission data have been normalized by dividing by the 
number of angstroms in the corresponding wavelength region.  
The middle and bottom panels show the continuum and 
emission data, respectively, as 1-$\sigma$ error bars 
folded on the orbital ephemeris of \citet{biro00}.
The solid curve(s) in each panel are sine 
function fits to the data with period fixed equal 
to the orbital period of DW UMa.  
\label{f-fuvlc}}
\end{figure}

In order to explore any rapid FUV variability in DW UMa, 
we repeated the extraction process described in Section \ref{s-mainobs}, 
but split each \fuse exposure into bins of 60 sec duration.  
This resulted in 44--56 spectra per exposure.  The individual 
spectra are of too low S/N to be useful in their dispersed 
form, but by summing the fluxes over selected wavelength 
regions, we can construct FUV light curves for DW UMa.  
The top panel of Figure \ref{f-fuvlc} shows the resultant 
light curves for continuum and emission line regions summed 
from the LiF2 spectra.  

\begin{figure}[tb]
\epsscale{0.80}
\plotone{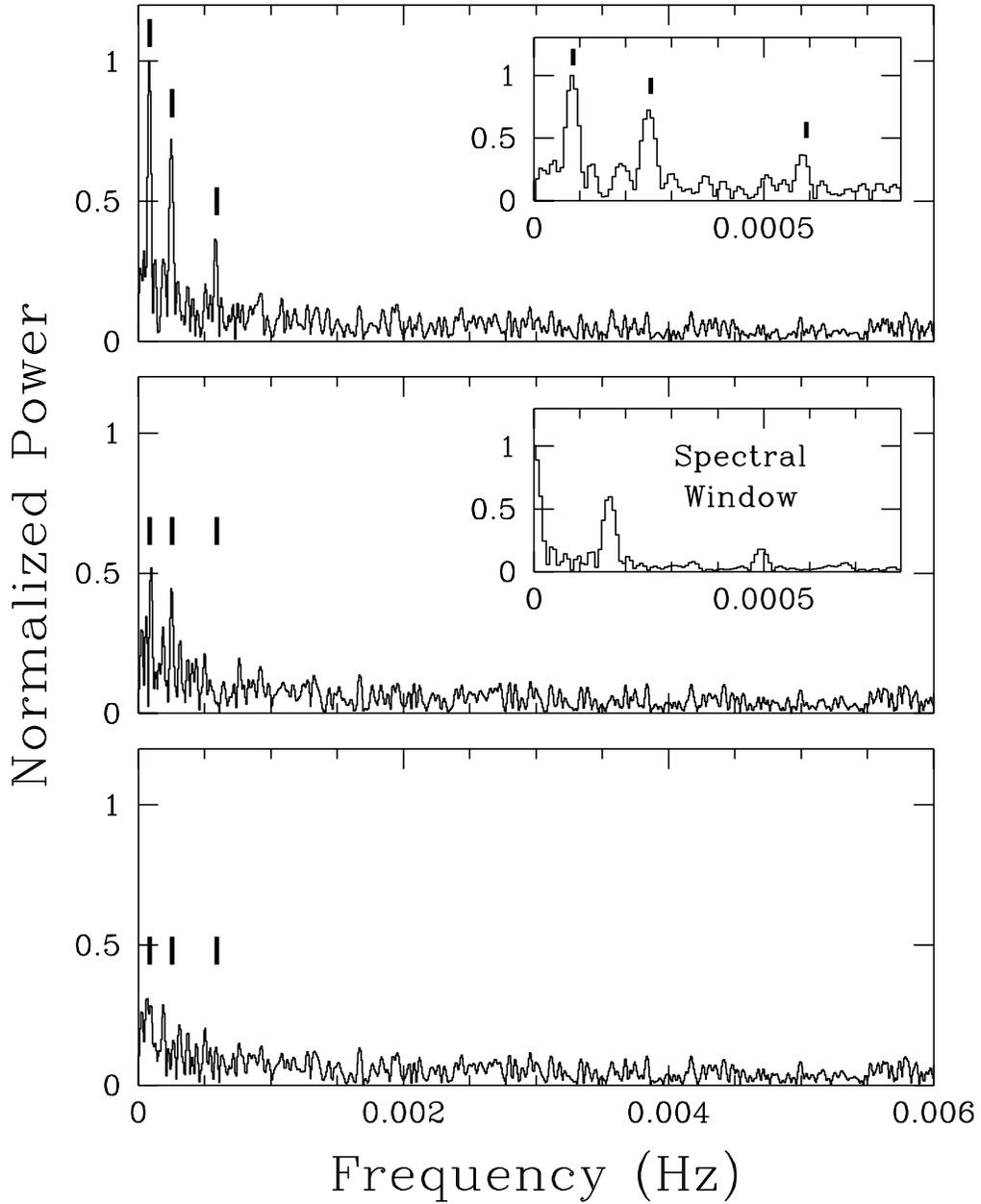}
\epsscale{1.00}
\figcaption{Power spectra calculated from the emission line data
shown in Figure \ref{f-fuvlc}.  The top panel shows the power 
spectrum of the original data.  The three marked peaks correspond 
(from low to high frequency) to the orbital period of DW UMa and 
aliases of $P_{\rm orb}/3$ and $P_{\rm orb}/7$.  The inset box 
in the top panel is an expanded view of the power spectrum at 
low frequencies.  The middle panel shows the power spectrum of 
the data after subtracting the best sinusoidal fit (with 
$P=P_{\rm orb}$).  The bottom panel shows the power spectrum of 
the data corresponding to the middle panel with an additional 
sinusoidal fit (with $P=P_{\rm orb}$) subtracted.  
Frequencies corresponding to the three peaks in the top panel are 
indicated in the other panels.  All three panels have been normalized 
to the maximum value in the top panel.  The inset box in the 
middle panel shows the spectral window.
\label{f-powspec}}
\end{figure}

We performed period searches on the light curves using a 
number of methods:\ power spectrum analysis including application 
of the {\sc CLEAN} algorithm \citep{roberts87}, 
phase-dispersion-minimization \citep{stel78}, and sine-wave 
fitting with period as a free parameter (using custom IDL code).  
The only periodicity that we found corresponds to the orbital 
period of DW UMa ($P_{\rm orb} = 11803$ sec, as refined from the 
ephemeris of \citealt{dhillon94} using an additional 14 optical 
eclipse timings obtained by \citealt{biro00}), along with two 
aliases of the $P_{\rm orb}$ signal.
No coherent rapid variability on shorter time scales was detected.  
The top panel of Figure \ref{f-powspec} shows the ``dirty''
(i.e., unCLEANed) power spectrum of the emission line light curve, 
with the orbital and alias peaks identified.
Because the periodic variability in the light curve is not 
strictly sinusoidal, subtracting the sine fit leaves a residual 
of the orbital modulation that is still the strongest signal in 
the power spectrum (see the middle panel of Figure \ref{f-powspec}).  
Another iteration of subtracting a sine fit from the data, however, 
removes almost all of the orbital signal from the power spectrum 
(see the bottom panel of Figure \ref{f-powspec}).  Yet, even with 
the modulation on $P_{\rm orb}$ removed, no other periodic signal 
is apparent in the power spectrum.

The middle and bottom panels of Figure \ref{f-fuvlc} show the 
continuum and emission line data, respectively, folded on the 
orbital period of DW UMa.
The modulation on $P_{\rm orb}$ is present in all three of the 
orbital periods of DW UMa spanned by the \fuse observations. 
The absolute amplitude of the modulation is larger in the emission 
line data, but the fractional semi-amplitude is comparable 
($\approx50$--60\% of the mean flux level) 
in both the emission line and continuum data.  
The phasings of the sine fits to the emission and continuum light curves are 
identical to within $\Delta\phi=\pm0.01$.
We also examined additional continuum and emission line 
(e.g., \ion{O}{6}, \ion{C}{3}) regions 
in the LiF2 spectra, as well as regions in the LiF1 spectra that 
either overlap with LiF2 regions or are unique to the LiF1 
wavelength coverage.  All regions show consistent characteristics 
and vary on $P_{\rm orb}$; the LiF2 data shown in 
Figure \ref{f-fuvlc} are representative of the entire data set.

In the optical, the light curve of DW UMa does not show a 
pre-eclipse hump caused by a bright spot where the accretion 
stream strikes the outer edge of the disk, nor any other periodic, short
time-scale variability other than the eclipse 
(e.g., \citealt{shafter88,dhillon94}).  
Our FUV light curve, on the other hand, peaks shortly before the 
eclipse ($\phi=0.93 \pm 0.01$ inferred from the sine fit shown in 
Figure \ref{f-fuvlc}) and has a minimum half an orbit later.  
This behavior is reminiscent of the orbital variability 
expected from irradiation-induced heating of the secondary star, 
but is out of phase with that mechanism (which should produce a 
maximum near $\phi=0.5$, when the secondary star is viewed face-on).  
The coincidence of the phase of the observed FUV maximum with that 
expected for the pre-eclipse hump suggests that whatever is producing the 
modulation in our FUV light curve is related to the accretion flow.

\subsection{Orbital-phase-resolved FUV Spectra}

\begin{figure}[tb]
\epsscale{0.80}
\plotone{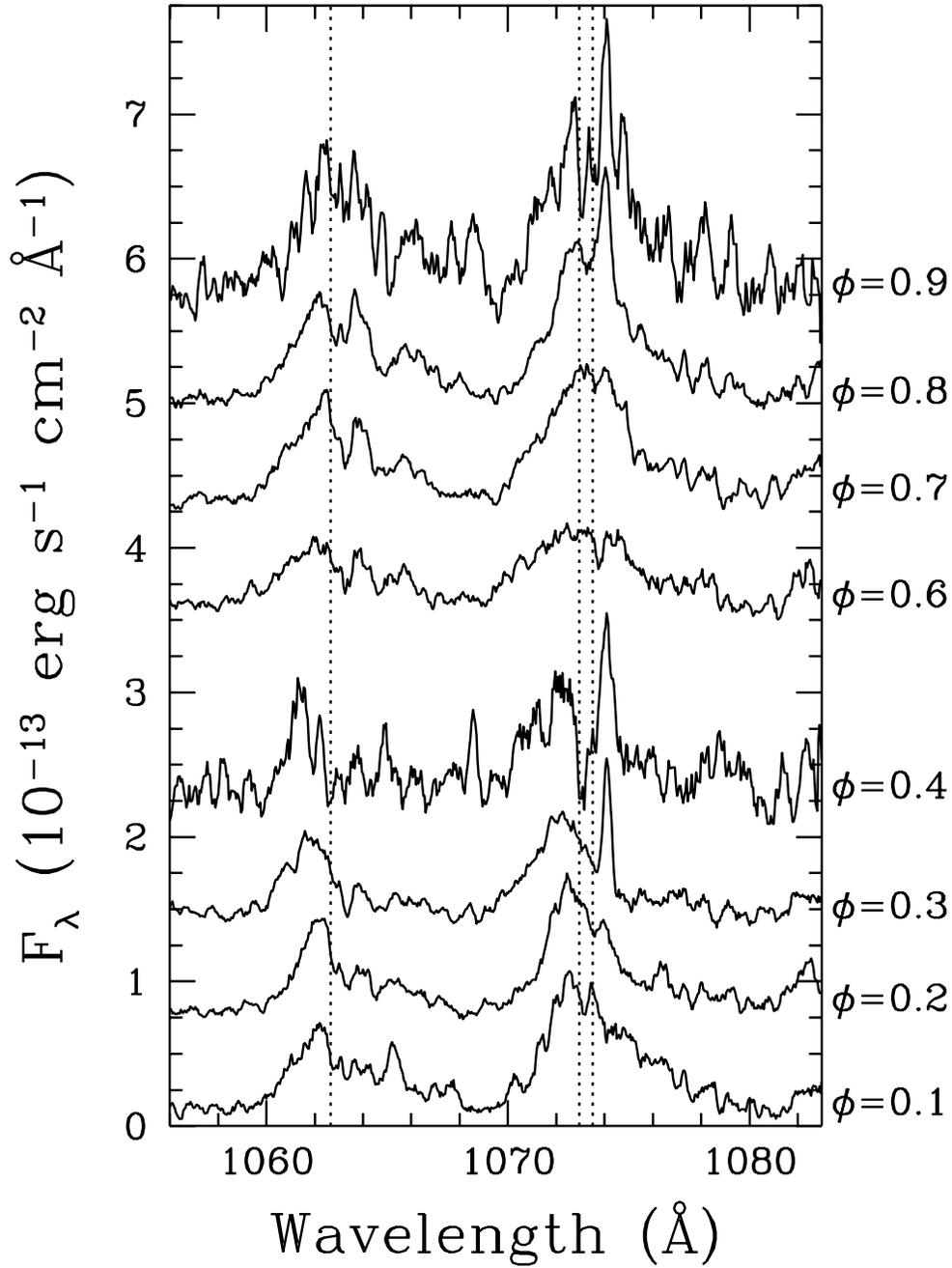}
\epsscale{1.00}
\figcaption{Orbital-phase-resolved \fuse spectra of DW UMa 
showing the region around the \ion{S}{4} multiplet (rest 
wavelengths are indicated by vertical dotted lines).  
Each spectrum above $\phi=0.1$ has been successively 
offset by $+0.7\times10^{-13}$ erg s$^{-1}$ cm$^{-2}$ \AA$^{-1}$ 
per $\Delta\phi=0.1$ (i.e., the $\phi=0.4$ and $\phi=0.6$ 
spectra are separated by twice this amount to account for 
the ``missing'' $\phi=0.5$ spectrum).  
The spectra have dispersions of 0.05 \AA\ pixel$^{-1}$ and 
have been boxcar-smoothed by 7 pixels.  The narrow emission component 
present at $\lambda\approx1074$ \AA\ in some of the spectra 
is an airglow feature.
\label{f-phispec1}}
\end{figure}

\begin{figure}[tb]
\epsscale{1.00}
\plotone{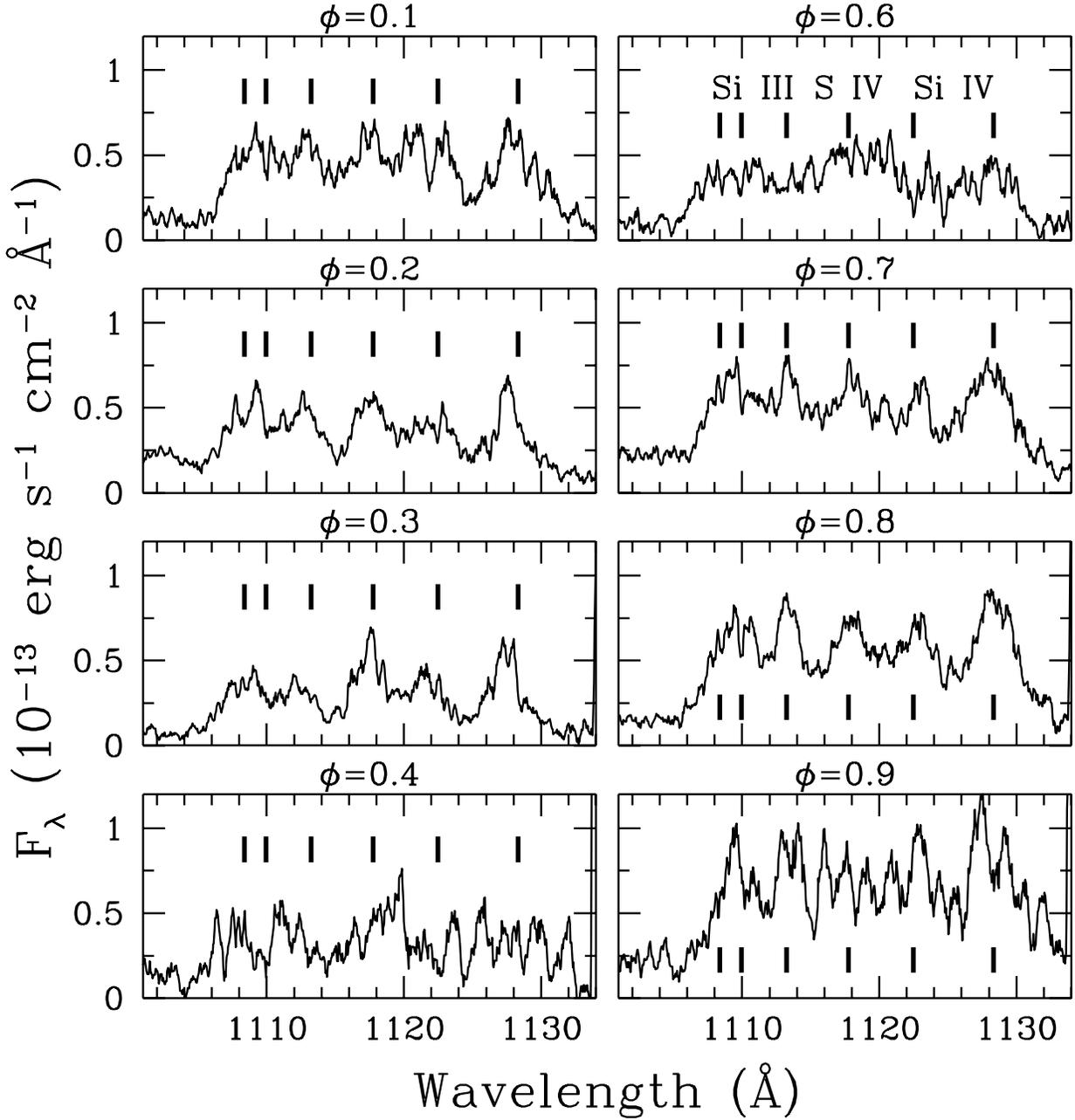}
\epsscale{1.00}
\figcaption{Orbital-phase-resolved \fuse spectra of DW UMa 
showing the region around the Si+S emission complex at 1105--1130 \AA.  
Vertical hashmarks show the rest wavelengths of the multiplet 
emission lines in each panel; the ions are identified in 
the $\phi=0.6$ panel.  The \ion{P}{5} $\lambda\lambda1118, 1128$ 
doublet is also present in this wavelength region; 
for example, see the tallest peaks in the $\phi=0.3$ panel.
Phases differ by half an 
orbit ($\Delta\phi=0.5$) between panels on the left and 
right in each row.  The spectra have dispersions of 
0.05 \AA\ pixel$^{-1}$ and have been boxcar-smoothed by 7 pixels.
\label{f-phispec2}}
\end{figure}

\begin{figure}[tb]
\epsscale{0.80}
\plotone{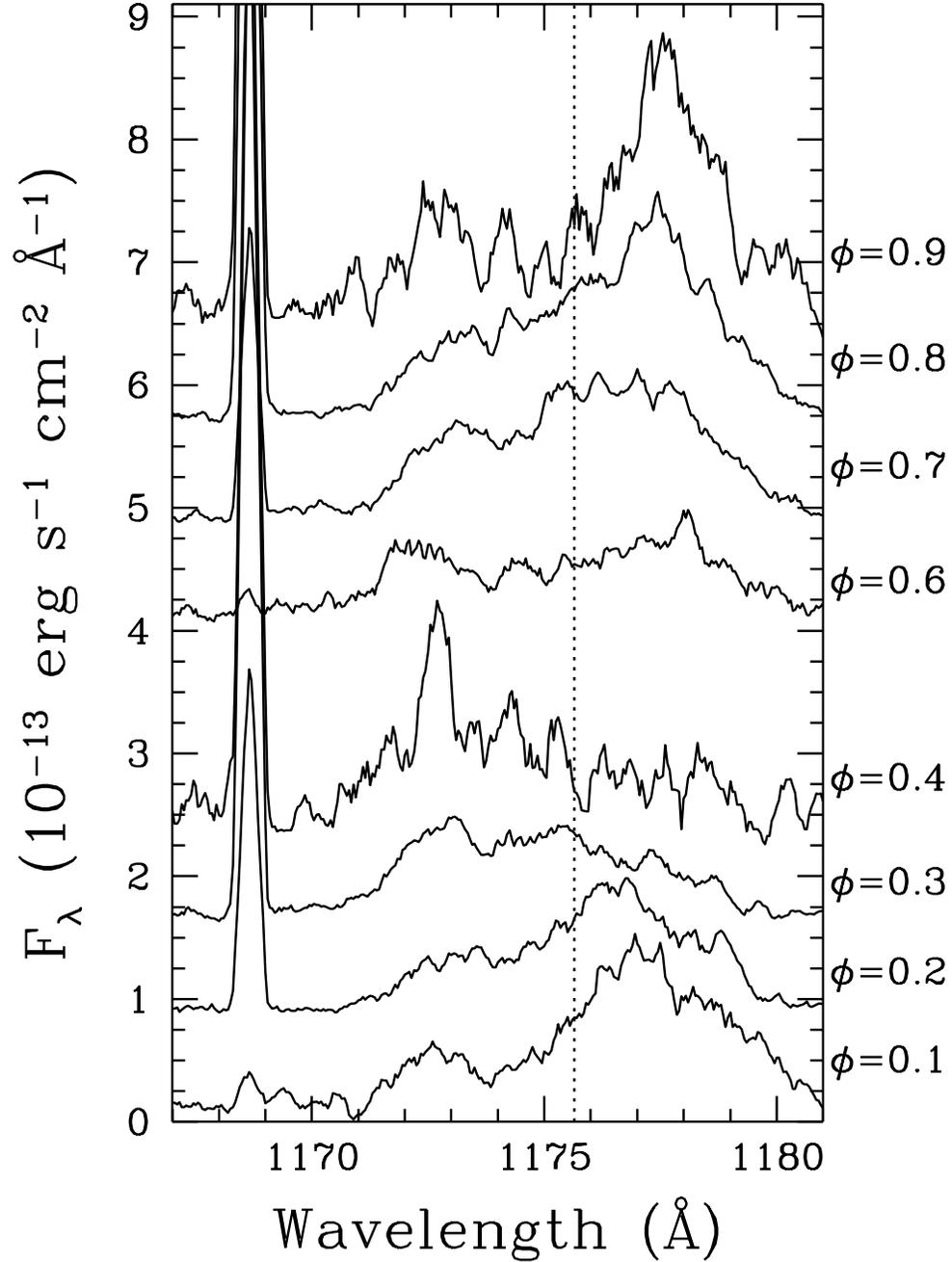}
\epsscale{1.00}
\figcaption{As in Figure \ref{f-phispec1}, but showing 
the region around the \ion{C}{3} multiplet.  A spacing increment 
of $+0.8\times10^{-13}$ erg s$^{-1}$ cm$^{-2}$ \AA$^{-1}$ 
per $\Delta\phi=0.1$ was used.
The strong, narrow emission component present at 
$\lambda\approx1168$--$1169$ \AA\ in some of the 
spectra is an airglow feature.
\label{f-phispec3}}
\end{figure}

We repeated the spectral extraction yet again, this time using 
the photometric orbital ephemeris of \citet{biro00} to split the 
data into eight spectra corresponding to orbital phase bins centered 
around $\phi=0.1$, $0.2$, $0.3$, $0.4$, $0.6$, $0.7$, $0.8$, and $0.9$.  
The phase bins around $\phi=0.0$ and $0.5$ were not sampled during 
the \fuse observations.  Table \ref{t-phispec} gives total exposure 
times for the phase-resolved spectra and lists which \fuse exposures 
contributed to each spectrum.  
Sections of the phase-resolved spectra containing prominent emission 
lines are shown in Figures \ref{f-phispec1}--\ref{f-phispec3}.  
There are obvious orbital-phase-dependent differences in the line 
profile shapes and velocity shifts, although none of the lines 
shows a clear ``S-wave'' signature (e.g., \citealt{honey87}).  
Perhaps the best case for an S-wave is in the \ion{C}{3} multiplet 
shown in Figure \ref{f-phispec3}, in which the highest point in 
the line profile shifts from the red side of the emission profile 
at $\phi=0.1$, to the blue side at $\phi=0.4$--$0.6$, then back to 
the red side at $\phi=0.9$.

\section{Discussion}

\subsection{Phenomenological Models for DW UMa}

Of particular interest in Figure \ref{f-phispec3} are the narrow 
peaks seen in the \ion{C}{3} emission profile in the 
$\phi=0.4$ (blue-shifted peak) and 
$\phi=0.9$ (red-shifted peak) phase bins.  
We verified that the peaks appear (with somewhat variable relative strengths)
in data from each of the \fuse exposures that contribute to the 
time-resolved spectra for the $\phi=0.4$ and $0.9$ phase bins.
The fact that these peaks are observed 
with opposite velocity shifts in spectra separated by half an orbit 
suggests that they originate from the same physical region 
in the CV.  
The separation and widths of the peaks imply a mean velocity of 
$\approx \pm650$ km s$^{-1}$ and velocity dispersion 
of $\approx 200$--$300$ km s$^{-1}$, respectively, for the FUV 
emitting region.
At the same time, the FUV light curve of DW UMa has a minimum 
at $\phi=0.43$ and a maximum at $\phi=0.93$.  
Here, we explore the relative merit of several phenomenological 
models for the structure of
DW UMa that might qualitatively account for this observed FUV behavior.

\subsubsection{Model 1:\ Accretion with ``Simple'' Stream Overflow}
\label{s-model1}

\begin{figure}[tb]
\epsscale{0.80}
\plotone{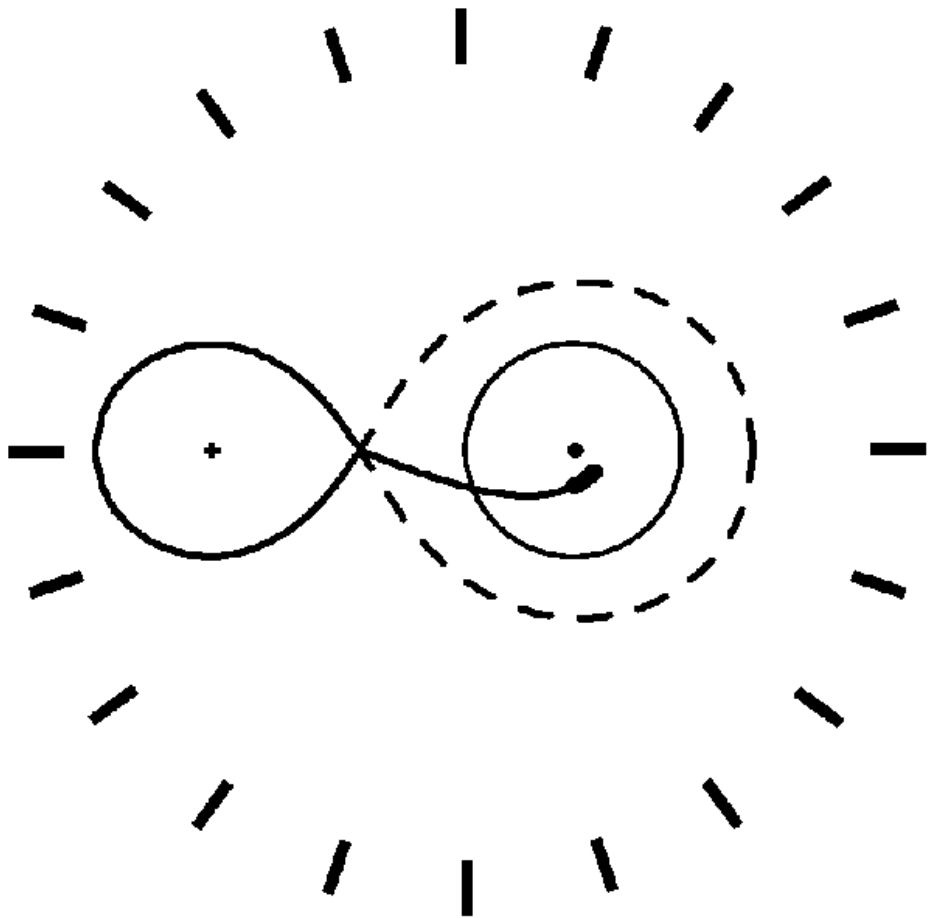}
\epsscale{1.00}
\figcaption{Schematic diagram of DW UMa for the ``simple'' 
accretion stream overflow model, depicting a face-on view 
of the Roche-lobe-filling secondary star (on the left) and 
the Roche lobe of the WD (dashed boundary on the right) 
containing the disk (large circle) and WD (small black dot).  
The accretion stream, which emerges from the L1 point, 
intersects the outer edge of the disk, and continues over 
the disk face.  The Roche lobes are to-scale using the 
component mass values determined by \citet{araujo03}.
The bold region at the terminus of the accretion stream 
is the emitting region discussed in Section \ref{s-model1}.
The hashmarks around the diagram mark lines of sight at 
orbital phase increments of $\Delta\phi=0.05$.  Phase 0.0 is at the 
9 o'clock position, and orbital phase increases clockwise 
(i.e., the CV rotates counter-clockwise relative to a 
fixed line-of-sight).
\label{f-diagram1}}
\end{figure}

Overflow of the accretion stream from its initial impact site 
with the disk edge and across the face of the disk 
(e.g., \citealt{lubow89}, \citealt{hessman99}) has been frequently 
proposed to play an important role in the origin of the SW Sex syndrome.
A schematic diagram of a CV in which the stream continues 
over the disk face is shown in Figure \ref{f-diagram1}.
This is similar to the diagrams shown by \citet{hell94} 
and \citet{hell96} in early attempts to explain the SW Sex stars.
At $\phi\approx0.9$, the terminus of the overflowing accretion stream 
is a kinematically plausible source for the red-shifted emission 
peak seen in DW UMa, since the stream trajectory 
(as well as the disk rotation) 
is directed approximately along 
this line-of-sight, in recession from the observer.  
Similarly, at $\phi=0.4$, this region is viewed from the opposite side, 
with a net matter flow toward the observer, thereby producing 
the blue-shifted emission peak.

In the context of the SW Sex stars, the terminus of the overflowing 
accretion stream in the inner disk has also been blamed for the 
distribution of emission regions in their Doppler tomograms, 
their radial velocity curve phase offsets, and the presence of 
the transient absorption feature.  However, in the case of 
DW UMa (and, by extension, possibly the other SW Sex stars also), 
this model has a fatal flaw.
\citet{knigge00} concluded that the disk in DW UMa is strongly flared,
and self-occults the WD and inner disk region.
This would make it difficult (perhaps even impossible, when the disk 
is oriented with the stream terminus on the side closest to us) 
to observe the stream 
terminus in the inner disk, especially in a high inclination CV.
This conclusion is supported by the \fuse spectra presented here, 
which show too little blue flux compared to accretion disk model 
spectra, indicating that (one way or another) the inner disk is 
not contributing to the spectrum.

\subsubsection{Model 2:\ Accretion with ``Complex'' Stream Overflow}
\label{s-model2}

\begin{figure}[tb]
\epsscale{0.80}
\plotone{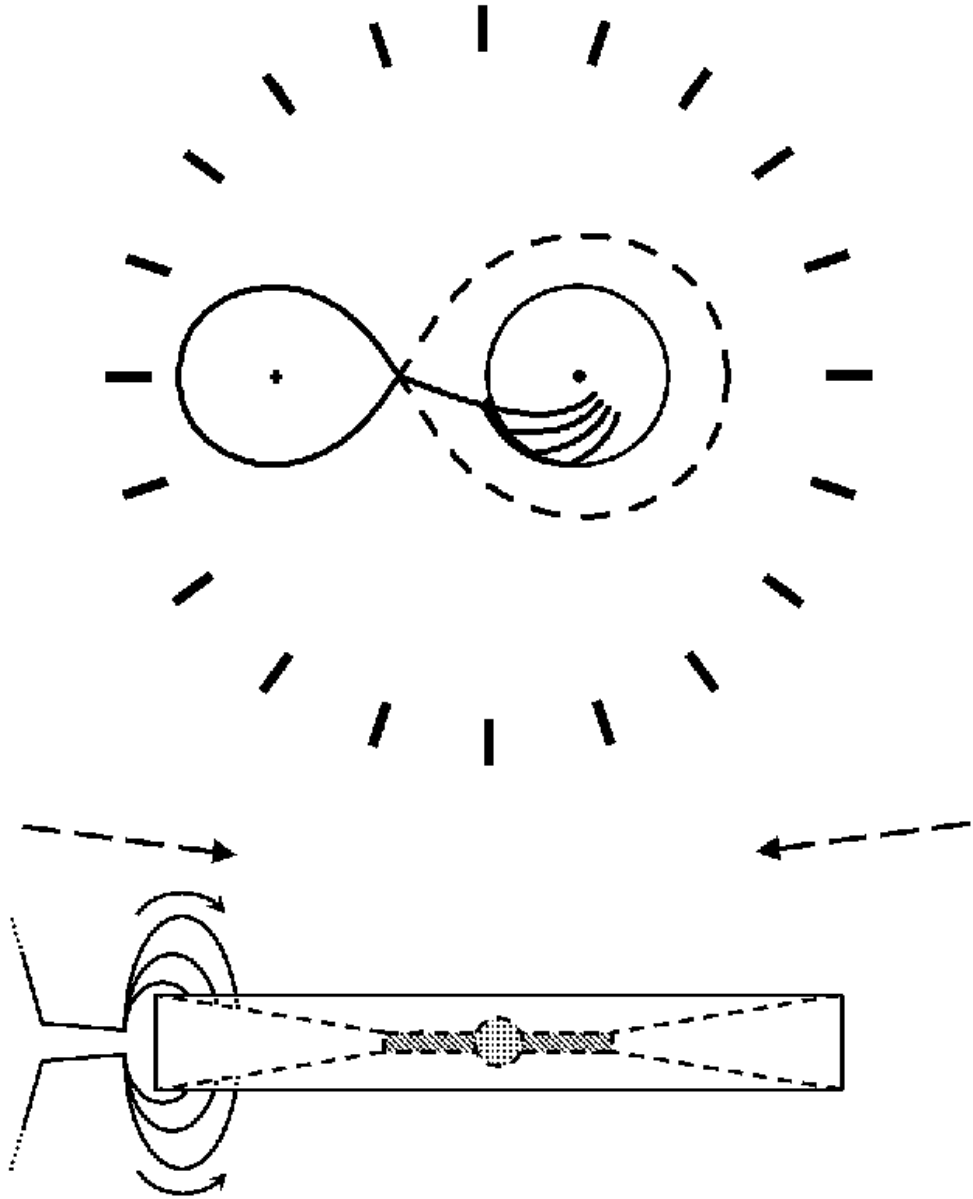}
\epsscale{1.00}
\figcaption{Schematic diagram of DW UMa for the ``complex'' 
accretion stream overflow model. The top sketch shows the 
face-on view (see description in the caption to 
Figure \ref{f-diagram1}), while the bottom sketch shows 
the edge-on view (at $\phi=0.75$, not to scale).  
In the bottom sketch, the flared disk rim (solid rectangle) 
is shown with an 
internal view of the disk radial profile (dashed lines), 
truncated inner disk (small cross-hatched rectangle), 
and WD (grey circle).  The two dashed arrows show 
lines-of-sight from Earth relative to the 
$i\approx82^{\circ}$ inclination of DW UMa.
\label{f-diagram2}}
\end{figure}

One way to overcome obscuration by a flared disk in a high 
inclination CV like DW UMa is to shift the emission region from 
the inner to outer disk.
Simulations of the stream-disk interaction in CVs 
(e.g., \citealt{arm96,arm98,kunze01}) 
have shown that the stream impact can result in an 
explosive ``splash'' that sends stream material along the edge 
of the disk (in the direction of disk rotation) and to 
substantial vertical height out of the disk plane.
Figure \ref{f-diagram2} shows a schematic diagram of a CV with 
complex stream overflow in which stream material splashes up against 
the disk edge, and is swept down the edge of the disk, 
resulting in a ``fan'' of overflowing material.  
A model of this type for the SW Sex stars has also been 
extensively developed by \citet{hoard98}, \citet{hoardetal98}, 
and \citet{hoard00}, so we will not reproduce those discussions 
in detail here.
As with Model 1, this model can account for the velocity-shift 
behavior of the \ion{C}{3} emission line peaks.  
More importantly, both the location of the stream flow near the 
disk edge, and its vertical extension above the plane of the disk, 
allow the emitting region to be seen at high inclination despite 
the flared disk.  
A possible problem with this model is that the velocity dispersion 
of emitting material in the fan of overflowing material might be 
expected to be larger than that inferred from the widths of the 
narrow line peak components seen in DW UMa.

This model also offers a possible explanation for the FUV light curve 
of DW UMa, if the overflowing material initially emits strongly in 
the FUV (when it is visible along the $\phi\approx0.9$ sightline) 
but not after it is flowing over the disk (when it is visible 
along the $\phi\approx0.4$ sightline).
Essentially, this is the same mechanism that produces a pre-eclipse 
hump in the light curves of CVs that have a bright spot where the 
accretion stream initially impacts the edge of the disk (but does 
not produce a hump in the light curve half an orbit later when 
the ``back'' side of the bright spot is visible).  
Unfortunately, this is difficult to reconcile with the lack of a 
bright spot in optical light curves of DW UMa, unless we speculate that the 
accretion stream impact/overflow site 
is uniformly hot enough that it emits only a negligible amount 
of luminosity in the optical wavelength range.  
(In principle, this is problematic for all three models discussed here, 
since they all involve stream overflow that might produce a bright spot; 
however, a prominent initial stream impact resulting in enhanced vertical 
structure on the disk rim plays a key role in Model 2.)

\subsubsection{Model 3:\ Stream-fed Accretion in an Intermediate Polar}
\label{s-model3}

The physical geometry of an IP (i.e., a truncated inner disk with 
magnetically-controlled accretion curtains through which material 
flows from the inner disk edge to the WD) could account for many of the 
peculiar observational properties that originally defined the 
SW Sex stars, if:\
(1) the single-peaked line profiles arise in a magnetic 
accretion curtain close to the WD; 
(2) the shallow eclipse of low excitation lines results because 
material following the field lines initially rises above the disk plane; and 
(3) self-absorption in the accretion curtain accounts for the 
transient absorption feature.  
Thus, it is little surprise that recently there has been an 
increasing trend in the relevant literature to invoke a 
magnetic scenario for the SW Sex stars.
Before we discuss such a model in application to DW UMa, and 
in the interest of providing a reference for future discussion, 
we summarize here key points of evidence 
culled from various literature sources that 
support a magnetic scenario for the SW Sex stars:

\begin{itemize}
\item{Perhaps the most blatant piece of evidence is that the 
observational properties of some CVs have caused them to be 
independently classified as both possible IPs/magnetic CVs and 
as possible SW Sex stars -- see the summary in Table \ref{t-ipsex}.}

\item{The detection (albeit somewhat marginal) of circular 
polarization in LS Pegasi and V795 Herculis \citep{gil01a,gil01b}
points to the presence of a magnetic WD.}

\item{``Flaring'' in optical emission lines on time scales of 
tens of minutes, which is suggestive of the asynchronous spin of 
the WD in an IP, is present in DW UMa, V533 Herculis, BT Mon, and LS Peg 
(\citealt{szkody01}, \citealt{gil02} and references therein). 
The circular polarization observed in LS Peg and V795 Her is 
also modulated on similar time scales.}

\item{\citet{patterson02} observed superhumps and 
kilosecond QPOs in optical light curves of numerous SW Sex 
stars that can be ascribed to the presence of a WD magnetic field.  
(Note:\ these authors suggest that the magnetic fields in SW Sex 
stars are comparable in strength to the highly magnetic polars, 
rather than being the weakest of the IPs, as is commonly asserted 
in other literature sources.)} 

\item{Hot spots in the disks of SW Sex stars that are revealed 
by eclipse mapping and/or Doppler tomography might be produced 
when the overflowing accretion stream encounters the WD magnetosphere; 
for example, as speculated by \citet{gil01b} for the observations 
of SW Sex by \citet{groot01}.}

\item{The typical Doppler tomograms of SW Sex stars 
(e.g., \citealt{kaitchuck94, hoard98}) might also be explained by 
a magnetic propellor mechanism in which some material in the 
overflowing accretion stream is ejected from the inner disk by 
the spinning WD magnetosphere \citep{horne99}.  (Note:\ this is 
a somewhat extreme variant of the magnetic scenario for SW Sex 
stars; to our knowledge, little additional work has been presented 
in support of it.)}

\item{Optical eclipse mapping studies of DW UMa \citep{biro00}, 
SW Sex \citep{groot01}, V1315 Aql, and other SW Sex stars 
(e.g., \citealt{rutten92}) show that their disk temperature 
profiles are best reproduced by assuming that the inner disk 
is suppressed or missing, as in the IPs.  
These results are somewhat ambiguous, as the temperature 
profiles could also be attributed to obscuration by the 
rim of a flared accretion disk.
\citet{knigge00} showed that the V-shaped eclipses seen in 
DW UMa and other SW Sex stars can be accounted for by the 
flared accretion disk.}

\item{The low inclination SW Sex stars V795 Her \citep{casares96}, 
LS Peg \citep{pais99,taylor99}, and V442 Oph \citep{hoard00} 
display Balmer emission line components extending to large velocity 
offsets from the line centers ($\Delta v \approx 1500$--$2000$ 
km s$^{-1}$), which produce prominent orbital S-waves in trailed spectra 
of these systems.  It has been suggested (e.g., \citealt{pais99}) 
that these components are present in the high inclination systems 
also, but at smaller velocity offsets such that they do not detach 
from the emission line cores and are, instead, visible as 
wings of the Balmer lines.  This behavior would require the 
emitting material responsible for the narrow components to have 
a vertical extent that allows a larger radial velocity to be seen 
at lower inclination.  While this is consistent with the expected 
structure for the magnetically-controlled accretion flow in a polar 
or IP, it also might be explained by a non-magnetic model involving 
emitting material located out of the disk plane (e.g., Model 2 
in Section \ref{s-model2}).}

\item{Several SW Sex stars (including DW UMa) are also 
VY Scl stars.  \citet{hameury02} have shown that the 
presence of a magnetic WD can suppress dwarf nova outbursts 
during the characteristic low states of the VY Scl stars 
(when the mass transfer rate is small enough that the disk 
instability mechanism is no longer quenched).  We note, however, 
that (as described by \citealt{hameury02}) the presence of a 
low mass ($M\lesssim0.4M_{\odot}$) and/or very hot 
($T\gtrsim40,000$ K) WD can also suppress dwarf nova outbursts 
during VY Scl low states.  The system parameters determined by 
\citet{araujo03} show that 
DW UMa does not satisfy the former condition 
($M\approx0.8M_{\odot}$), but it does satisfy the latter 
($T=50,000$ K).  Thus, a magnetic field is not necessarily 
required to suppress dwarf nova outbursts in DW UMa or other 
SW Sex stars with hot WDs.  In addition, \citet{hameury02} 
do not seem to have accounted for the fact that hotter WDs 
are larger \citep{koester86}, which (according to Equation 3 
in \citealt{hameury02}) would cause dwarf nova outbursts to 
be suppressed at a lower WD temperature\footnote{A more 
fundamental issue is that the low states of SW Sex stars may 
not have been well-enough observed to rule out the presence 
of dwarf nova outbursts.  For example, the long-term optical 
light curve of DW UMa presented by \citet{honeycutt93} has 
only sparse sampling of the low state.  The light curve shows 
several isolated data points during low states that are brighter 
by a magnitude or more compared to the typical low state 
brightness.  These might correspond to poorly sampled 
outbursts.}.}
\end{itemize}

The lack of rapid coherent modulations (period $\lesssim$ a few kiloseconds) 
in our FUV light curve 
of DW UMa is an important factor that argues against a standard IP geometry  
for this CV.  Such modulations are seen in known IPs across 
a wide wavelength range, and are the hallmarks of 
the asynchronous spin of the WD \citep{patterson94}.  
We are, perhaps, saved in this case because the WD in DW UMa 
is hidden by the disk in the high accretion state.  
Even so, we might expect to detect the spin period via 
reprocessing of light in other parts of the CV that {\em are\/} 
visible to us.  As mentioned above, emission line ``flaring'' with a
characteristic time scale on the order of several tens of minutes, which
may be linked to the spin of a magnetic WD, has
been tentatively identified in DW UMa in the optical. However, 
this does not help to explain the FUV behavior of this CV, which shows
interesting spectral and photometric variability modulated only 
on the orbital period, instead of any putative, shorter WD spin period.

\begin{figure}[tb]
\epsscale{0.80}
\plotone{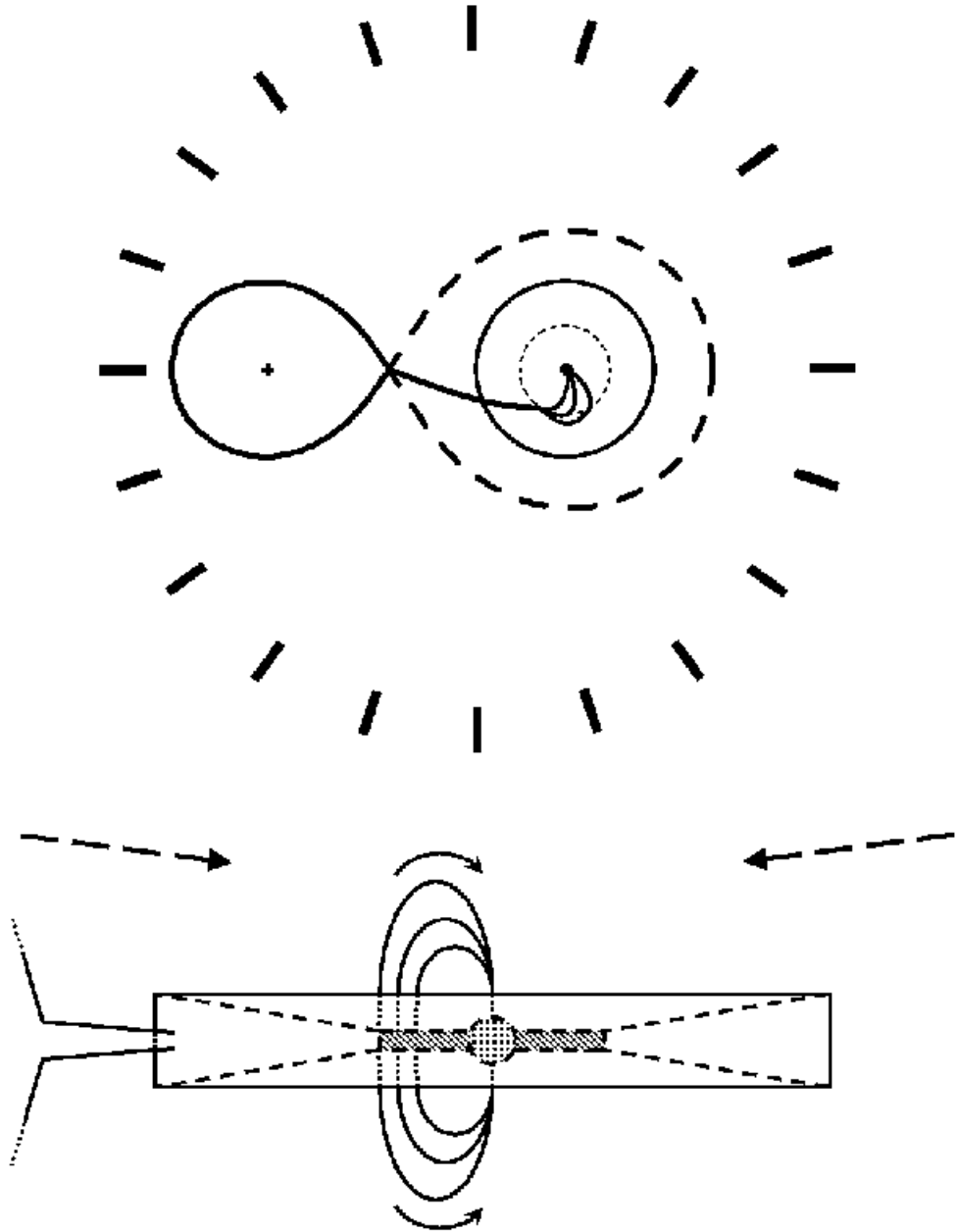}
\epsscale{1.00}
\figcaption{Schematic diagram of DW UMa for the stream-fed 
IP model (see description in the caption to Figure \ref{f-diagram2}). 
The inner edge of the truncated disk is shown as a dotted circle 
in the top sketch.
\label{f-diagram3}}
\end{figure}

An alternative to the standard, accretion-curtain model for IPs is
a stream-fed scenario, in which matter in an overflowing accretion 
stream encounters the WD magnetosphere at the inner edge of the 
truncated disk (e.g., \citealt{hell93}, \citealt{norton97}, 
\citealt{ferrario99}). 
The stream material is then entrained onto the field lines 
and funneled onto the WD.  
\citet{groot01} suggested that SW Sex itself shows evidence 
for a shocked hot spot in the inner disk that could occur when 
the overflowing stream hits the spinning magnetosphere.
In this model (which is depicted schematically in 
Figure \ref{f-diagram3}), the FUV-emitting region is moved back toward 
the inner disk, but the vertical extension of the accretion 
funnel above the plane of the disk allows us to observe it 
despite the obscuring rim of the flared disk.
In principle, there should be two 
symmetric (more or less) funnels above and below the disk;
geometric effects could cause only one of these to be visible in
some systems.
Hereafter, we assume a basic structure for the 
accretion funnel (e.g., \citealt{fw99}) 
in which it is broad at the base (where it emanates from 
the stream overflow at the inner edge of the truncated disk)
and narrow at the tip (where it connects 
to the pole of the WD); both temperature and density increase 
from base to tip.  
In general, the bottom side of the 
funnel (facing the WD+disk) will be hotter than the top side, 
because the former 
is irradiatively heated by the WD; the front side of the funnel 
(facing the WD)
will be hotter than the back side because of both irradiative 
heating from the WD and the intrinsic temperature profile of the funnel.

Model 3 is different from the standard IP scenario, in that the 
location of the accretion funnel is tied to the overflowing 
stream trajectory in the disk, so it does not rotate with 
the WD like the accretion curtains do.  
If the magnetic axis of the WD is not severely misaligned with 
its spin axis, then this results in a persistent accretion 
funnel extending from the terminus of the overflowing stream 
onto the polar region of the WD.  
The spinning accretion curtains of a standard IP may or 
may not also be present; however, the lack of a short spin period 
modulation in our \fuse data implies that if they are present in 
DW UMa, then they are either not visible or do not contribute 
significantly in the FUV.
If the magnetic and spin axes of the WD are misaligned, causing the 
tip of the funnel to ``follow'' the spin of the magnetic pole, 
then this could introduce a modulation on $P_{\rm spin}$ (or more 
likely on the beat period between $P_{\rm spin}$ and $P_{\rm orb}$) 
into the light curves, whose amplitude increases as the misalignment 
increases.
Alternatively, the ``active'' funnel might switch between above 
and below the disk as the WD spins, to accomodate accretion onto 
the more geometrically favorable pole.  Again, this could result 
in a modulation on $P_{\rm spin}$.
Clearly, the lack of detectable WD spin modulations, although 
mitigated somewhat in Model 3 compared to a standard IP model, 
is a hindrance to magnetic models for the SW Sex stars.

In DW UMa, near $\phi=0.93$ (before eclipse when the secondary 
star blocks our view of the disk), we see a maximum in the FUV 
light curve caused by the luminosity of the back side of the 
accretion funnel.  A red-shifted peak is produced in the \ion{C}{3} 
emission line because from that perspective, material in the 
funnel is flowing away from us with a relatively small velocity dispersion.  
At quadrature phases, we see only a moderate brightness in the 
FUV light curve due to the smaller projected area of the ``side'' 
of the accretion funnel.  At the same time, there is little or 
no evidence for a preferred velocity shift in the line profile 
because the matter flow is mostly tangential to this line-of-sight.   
This is the perspective looking into the page at the bottom sketch in 
Figure \ref{f-diagram3}.
Near $\phi=0.4$, we see primarily the front side of the 
accretion funnel.  This produces a blue-shifted emission 
peak because material in the funnel is moving towards us 
along this line-of-sight.  
The FUV light curve displays a minimum at this phase because 
less FUV luminosity in the \fuse sensitivity band 
(i.e., $\lambda \approx 1000$--$1180$ \AA\ for the light curves 
shown here) is emitted from the hotter front side of the 
funnel than from the cooler back side. 
The front side has a smaller projected surface area than the 
back side, so we still see some of the back side; however, 
this is the bottom of the back side, which is also hotter 
than the top of the back side (which is viewed at $\phi=0.93$).
This difference between the temperatures and emissive properties 
of the front and back sides of the funnel is supported by the 
difference in the profile shapes of the corresponding emission 
line peaks:\ the red-shifted, back-side peak is stronger and 
broader than the blue-shifted, front-side peak.

\section{Conclusions}

We have presented the first FUV spectra of the archetype SW Sex star, 
DW UMa, which are dominated by broad emission lines of a variety of ions of 
H, C, N, O, Si, and S.  The emission lines vary in flux, profile shape, 
and velocity shift during the orbit of the CV; in particular, the 
\ion{C}{3} emission complex at 1178 \AA\ 
shows S-wave-like profile variations as a function of orbital phase.
The relatively weak continuum in the time-averaged spectrum is difficult 
to model in terms of a steady state accretion disk.  Although there are 
no coherent, short time scale periodicities ($P\lesssim$ a few kiloseconds) 
in the FUV light curves of DW UMa, both the continuum and line fluxes 
vary on the orbital phase, with a maximum at $\phi=0.93$ (just prior to 
the inferior conjunction of the secondary star -- i.e., the eclipse) 
and a minimum half an orbit later.

It is clear from our \fuse observations that DW UMa displays 
substantial activity in the FUV wavelength regime.  
The hot inner region of this high-inclination CV, close to the WD 
(i.e., the canonical origin for high energy luminosity in 
disk-fed CVs), is believed to be self-occulted by a flared 
accretion disk rim; it may even be missing altogether, truncated 
by a spinning WD magnetosphere. 
This effectively rules out a ``simple'' stream overflow model 
(at least for high inclination SW Sex stars) in which the 
inner disk must be visible at all phases 
to account for their observational properties.
Our \fuse data did not yield direct evidence linking 
DW UMa to the IP class of magnetic CVs.  
Nonetheless, the stream-fed
IP scenario described in Section \ref{s-model3} offers at least 
a qualitative explanation for the FUV behavior of DW UMa.  
Observational evidence continues to mount that 
magnetic fields may play a key role in causing the SW Sex 
syndrome that has infected many novalike CVs.  
The SW Sex stars may indeed be the IPs with the highest mass accretion 
rates and/or weakest magnetic fields.  
That said, the IP model is certainly not {\em demanded\/} by 
our \fuse data.  
The lack of rapid coherent modulations 
in our FUV light curves 
of DW UMa, corresponding to the WD spin, casts doubt on the IP scenario.
The ``complex'' accretion stream overflow model (see 
Section \ref{s-model2}) can, arguably, offer an equally 
satisfactory explanation, especially when considered in conjunction 
with the presence of a flared, self-occulting accretion disk in DW UMa.
We note, however, that these two models are not necessarily 
mutually exclusive and could operate simultaneously, 
with each mechanism contributing differently in different SW Sex stars. 
In particular, the model that (eventually) successfully explains 
the SW Sex stars (including the several known low-inclination 
systems) cannot rely solely on the presence of a flared disk, which 
is most explanatory in high inclination systems.
Compelling evidence for the magnetic nature of the SW Sex stars 
would include reliable and repeatable detections, for many 
SW Sex stars, of both polarization and coherent modulations 
that can be securely linked to WD spin.

\acknowledgements

D.\ W.\ H.\ thanks Stephan McCandliss for a helpful discussion 
about the characteristics of molecular hydrogen in \fuse spectra 
and for an introduction to his {\sc H2ools} package.  
This work was supported by NASA \fuse grant NAG 5-10246.  
The research described in this paper 
was carried out, in part, at the Jet Propulsion Laboratory, 
California Institute of Technology, and was sponsored by the 
National Aeronautics and Space Administration.
We made use of the SIMBAD database, operated at CDS, Strasbourg, France.


\input{tab1}
\input{tab2}
\input{tab3}
\input{tab4}


\end{document}

%% file: tab1.tex
\begin{deluxetable}{cccc}
\tablenum{1}
\tablewidth{0pt}
\tablecaption{{\em FUSE} Observation Log \label{t-fuselog}}
\tablehead{
\colhead{Exposure} &
\colhead{Start Time} &
\colhead{Total Exposure} &
\colhead{$\phi$\tablenotemark{a}} \\
\colhead{\#} &
\colhead{(HJD$-$2450000)} &
\colhead{(s)} &
\colhead{ } 
}
\startdata
 1 & 2221.260217 & 3256 & 0.059 \\
 2 & 2221.329574 & 3330 & 0.567 \\
 3 & 2221.398896 & 3332 & 1.074 \\
 4 & 2221.469028 & 3264 & 1.587 \\
 5 & 2221.541579 & 2988 & 2.119 \\
 6 & 2221.613471 & 2668 & 2.645 \\
\enddata 
\tablenotetext{a}{Orbital phase at start of exposure 
from the ephemeris of \citet{biro00}.}
\end{deluxetable}

%% file: tab2.tex
\begin{deluxetable}{lccc}
\tablenum{2}
\tablewidth{0pt}
\tablecaption{Selected Lines in the FUV Spectrum of DW UMa\label{t-lines}}
\tablehead{
\colhead{ } &
\colhead{ } &
\multicolumn{2}{c}{Gaussian Fit} \\
\colhead{Line} &
\colhead{EW} &
\colhead{EW} &
\colhead{FWHM} \\
\colhead{ } &
\colhead{(\AA)} &
\colhead{(\AA)} &
\colhead{(km s$^{-1}$)}
}
\startdata
\ion{H}{1} $\lambda919.35$\tablenotemark{a}  & 0.29 & \nodata & \nodata \\
\ion{H}{1} $\lambda920.96$\tablenotemark{a}  & 0.31 & \nodata & \nodata \\
\ion{H}{1} $\lambda923.15$\tablenotemark{a}  & 0.33 & \nodata & \nodata \\
\ion{H}{1} $\lambda926.23$\tablenotemark{a}  & 0.26 & \nodata & \nodata \\
\ion{S}{6} $\lambda944.52$                   & 11.5 & 12.7    & 815 \\
\ion{S}{3} $\lambda1012.50$                  & $\approx2$ & 2.0 & 730 \\
\ion{S}{3} $\lambda1015.6$\tablenotemark{b}  & 2.7  & 3.0 & 575 \\
\ion{S}{3} $\lambda1021.2$\tablenotemark{b}  & $\approx3$ & 3.7 & 795 \\
\ion{H}{1} $\lambda1025.72$\tablenotemark{c} & $\approx32$ & 36.3 & 1000 \\
\ion{O}{6} $\lambda1031.93$                  & 47.3 & 45.1 & 1245 \\
\ion{O}{6} $\lambda1037.62$\tablenotemark{d} & 23.6 & 27.5 & 1075 \\
\ion{S}{4} $\lambda1062.66$\tablenotemark{{\rm d,e,f}} \hspace*{2em}  & 21.7 & 20.9 & 1250 \\
\ion{S}{4} $\lambda1073.2$\tablenotemark{{\rm b,d,e,g}} & 34.7 & 29.3 & 1120 \\
\ion{Si}{3} $\lambda1142.7$\tablenotemark{b} & 10.5 & 11.9 & 1585 
\enddata 
\tablenotetext{a}{ISM absorption line.}
\tablenotetext{b}{Blend of multiplet transitions.}
\tablenotetext{c}{Blended airglow line is masked off.}
\tablenotetext{d}{Affected by interstellar absorption.}
\tablenotetext{e}{A Gaussian function is a poor fit to this line profile.}
\tablenotetext{f}{Possibly blended with \ion{Si}{4} $\lambda1066.63$.}
\tablenotetext{g}{Possibly blended with \ion{S}{3} $\lambda1077.16$.}
\end{deluxetable}

%% file: tab3.tex
\begin{deluxetable}{ccc}
\tablenum{3}
\tablewidth{0pt}
\tablecaption{Orbital-phase-resolved Spectra \label{t-phispec}}
\tablehead{
\colhead{Orbital Phase} &
\colhead{Data from} &
\colhead{Total} \\
\colhead{Bin Center\tablenotemark{a}} &
\colhead{Exposures} &
\colhead{Exposure} \\
\colhead{ } &
\colhead{ } &
\colhead{(s)}
}
\startdata
0.1 & 1, 3, 5 & 2340 \\
0.2 & 1, 3, 5 & 3540 \\
0.3 & 1, 3, 5 & 3360 \\
0.4 & 3, 5 & \phn330 \\
0.6 & 2, 4, 6 & 1780 \\
0.7 & 2, 4, 6 & 3540 \\
0.8 & 2, 4, 6 & 3525 \\
0.9 & 4, 6 & \phn410 \\
\enddata 
\tablenotetext{a}{Bin widths are $\Delta\phi=0.1$.}
\end{deluxetable}

%% file: tab4.tex
\begin{deluxetable}{lll}
\tabletypesize{\small}
\tablenum{4}
\tablewidth{0pt}
\tablecaption{CVs Classified as Both IPs/Magnetic CVs and SW Sex Stars \label{t-ipsex}}
\tablehead{
\colhead{CV Name} &
\colhead{IP Classification\tablenotemark{a}} &
\colhead{SW Sex star Classifiation}
}
\startdata
TT Arietis    & \citet{patterson94} & \citet{patterson02} \\ \\

V533 Herculis & \citet{patterson94} & \citet{gil02} \\
              & \citet{patterson79} & \citet{thor00} \\ \\

V795 Herculis & \citet{patterson94} & \citet{dick97}\tablenotemark{b} \\
              & \citet{zhang91}     & \citet{casares96}\tablenotemark{b} \\
              & \citet{shafter90}   &  \\ \\

EX Hydrae     & \citet{patterson94} & \citet{hell00} \\
              & \citet{kaitchuck87} & \\
              & \citet{vogt80}      & \\ \\

BT Monocerotis & \citet{white96}    & \citet{smith98} \\ \\

V348 Puppis   & \citet{tuohy90}\tablenotemark{c} & \citet{gil01c} 
\enddata 
\tablenotetext{a}{We have only listed here those CVs that were considered as possible magnetic systems {\em before} they were considered as possible SW Sex stars.  With the advent, circa 1996, of the IP-like model for SW Sex stars, there is no shortage of CVs that have been recently considered simultaneously as possible IPs and SW Sex stars.}
\tablenotetext{b}{These authors suggest a possible magnetic scenario for V795 Her as an SW Sex star.  \citet{rosen95,rosen98} argue against, but do not rule out, an IP model for V795 Her.}
\tablenotetext{c}{\citet{rosen94} conclude that ``...{\em ROSAT\/} observations of [V348 Pup] do not provide any conclusive evidence in favor of the IP classification...''}
\end{deluxetable}